\documentclass[a4paper,11pt]{article}
\usepackage{jcappub} 
\usepackage{lineno}
\usepackage{float}
\usepackage{subfigure}
\usepackage{xcolor}
\usepackage{amsmath}
\usepackage[normalem]{ulem}

\newcommand{\Ari}[1]{\textcolor{black}{#1}}

\title{\boldmath Comparative of light propagation in Born-Infeld, Euler-Heisenberg and ModMax nonlinear electrodynamics}

 \author[a,b]{Elda Guzman-Herrera,}
 \author[a]{Ariadna Montiel\note{Corresponding author.}}
 \author[a]{and Nora Breton}
 \affiliation[a]{Physics Department, Centro de Investigaci\'{o}n y de Estudios Avanzados del Instituto Polit\'ecnico Nacional (Cinvestav), PO. Box 14-740, Mexico City, Mexico}
 \affiliation[b]{Instituto de Física e Química, Universidade Federal de Itajubá, Itajubá, Minas Gerais 37500-903, Brasil}

\emailAdd{ariadna.montiel@cinvestav.mx}

\abstract{We compare light propagation through an intense electromagnetic background as described by three different nonlinear electrodynamics: Born-Infeld (BI), Euler-Heisenberg (EH), and Modified Maxwell (MM). We use the concept of effective metric to determine the phase velocities of a propagating wave from the BI and EH nonlinear electrodynamics and use them to set constraints on the MM nonlinear parameter. In a second part of the paper, we  consider the black hole solutions of the nonlinear electrodynamics coupled with General Relativity  and  determine  the shadows cast by the static black holes. Confronting the  observations of the shadows of M87$^*$ and Sagittarius A$^*$  with our theoretical results allows us to set restrictions on the nonlinear parameters.}

\begin{document}
\maketitle
\flushbottom

\section{\label{sec:level1}Introduction}

Nonlinear electromagnetic effects become significant when the electromagnetic field strengths approach the critical fields $E_{cr}\approx m^{2}_{e}c^{3}/(e\hbar)\approx 10^{18}$ Volt/m or $B_{cr}\approx 10^{9}$ Tesla; $B_{cr}$ represents the field at which the cyclotron energy equals $m_{e}c^2$, where $m_{e}$ is the mass of the electron, and it defines the field scale at which the external field becomes significant on quantum processes; it is also known as Schwinger critical field and it marks out the threshold of nonlinear QED scale \cite{Dunne2012}. Moreover, anyone familiar with quantum field theory is led to the conclusion that a classical phenomenological theory that pretends a description of polarization effects (light-light interaction or pair production in vacuum excited by an electric field) has to be nonlinear.


A strong magnetic field is a relatively common occurrence in the universe. For example, pulsars ($B \sim 10^{12} $G) \cite{Denisov2014EffectsON,Abishev:2016xxa} and magnetars ($\sim 10^9$ to $10^{11}$ Tesla, $ \sim 10^{13}$ to $10^{15}$ G) \cite{Kaspi:2017fwg,PhysRevD.71.063002,MosqueraCuesta:2003dh}, with 30 confirmed cases so far \cite{2014ApJS..212....6O}, exhibit intense magnetic fields where various nonlinear electrodynamic effects, such as vacuum birefringence, occur. Neutron stars can possess magnetic fields in the range of $10^6-10^9$ Tesla and processes such as photon splitting and pair conversion are expected to occur in their vicinity \cite{Baring2008, Mignani2017}. Some $\gamma$-ray pulsars are characterized by magnetic fields near the Schwinger critical field \cite{Denisov2017}. It is known that to achieve the measurement of the \Ari{nonlinear electrodynamics (NLED)} effects, instruments with large area to measure polarization are needed \cite{Denisov2017}. 

On the other hand, the PVLAS experiment aimed to observe electromagnetic vacuum birefringence but did not achieve the predicted value from QED. Even so, it did establish the current best limit on vacuum birefringence in the presence of an external magnetic field ($B_{ext}=2.5$ Tesla) \cite{Zavattini2008, Zavattini2012, Ejlli:2020yhk}. A new PVLAS experiment is expected to start soon, using more powerful magnets, to reach the QED value of birefringence \cite{Ejlli:2020yhk}. The BFRT Collaboration \cite{Cameron1993},  the BMV group \cite{Cadene:2013bva} as well as Hercules\cite{Tommasini:2009nh,Karbstein:2021otv}, attempted to test vacuum birefringence without success too. Due to this, new proposals have been made to establish boundaries for detecting vacuum birefringence\cite{Ahmadiniaz2020}. In \cite{fouche2016limits}, a comprehensive framework for conducting experiments to validate QED predictions has been proposed, thus providing an opportunity to test various NLED theories.

The other phenomena that could be useful to measure or set bounds on NLED is the analysis of photon-photon scattering or photon splitting. Light-by-light interactions can be studied using heavy-ion collisions, these intense EM fields can be treated as a beam of quasi-real photons and light-by-light scattering has been measured in $Pb+ Pb$ collisions at the Large Hadron Collider (LHC) \cite{Atlas2017, Enterria2013}. A larger study in experiments using intense laser and energetic particle beams focusing in QED is presented in \cite{Fedotov2023} and a more recent review of photon-photon scattering at LHC can be consulted in \cite{Schoeffel2021}. Nonetheless, to reach QED predictions there is needed a stronger magnetic field source, that currently is only achievable in astrophysical objects.

The nonlinear electromagnetic (NLEM) interactions can be described in a classical way using an effective Lagrangian that  depends nonlinearly on the two Lorentz and gauge invariants, $F$ and $G$, of the Faraday tensor $F_{\mu \lambda}=\partial_{\mu}A_{\lambda}-\partial_{\lambda}A_{\mu}$,

\begin{equation}
F=F^{\mu \lambda}F_{\mu \lambda}= 2(B^2-E^2), \quad G=F^{*\mu\lambda}F_{\mu\lambda}= -4 \vec{B} \cdot \vec{E},
\label{invariants}
\end{equation} 
where $F^{*\mu\lambda}=\frac{1}{2}\epsilon^{\mu\lambda\alpha\beta}F_{\alpha\beta} $.  

Maxwell's electrodynamics cannot describe nonlinear interactions between two electromagnetic waves, for example, the interaction light-by-light; these phenomena are well known in QED, and considering classical NLED it is possible to describe them \cite{tzenov2020dispersion}. 
Several NLED proposals exist to describe such nonlinear effects, but three of them stand out:  Born-Infeld (BI), Euler-Heisenberg (EH), and ModMax theories.

In 1934, Born and Infeld \cite{BI1934} presented a theory with nonlinear corrections to Maxwell electrodynamics with the aim of solving the singularity of the electric field self-energy of charged particles;  they propose the existence of a maximum attainable electromagnetic field, denoted by the BI parameter $b$, with a magnitude $b= e / r_0^2 = 10^{20}$ Volt/m, where $r_0$ is the classical electron radius. The interpretation of this absolute field is also that the classical self-energy of the electron is equal to its mass-energy at rest \cite{fouche2016limits}. 
The Lagrangian for this NLED is established as 
\begin{equation}
    L_{BI}=-4b^2 \left(1-\sqrt{1+\frac{F}{2b^2}-\frac{G^2}{16b^4}}\right).
    \label{LBI}
\end{equation}

This classical theory effectively models vacuum polarization as a material medium.  
From the viewpoint of QED, the BI modifications of the Maxwell theory result in a change in the photon-photon scattering signal. This is why it has been proposed that experiments that measure the QED process could constrain the maximum field $b$ of the BI theory \cite{Fedotov2023}. Another interesting feature is that it presents neither birefringence nor shock waves; the absence of birefringence leads to the idea that vacuum birefringence experiments could set bounds on the BI parameter $b$  \cite{Fedotov2023, Rebhan2017, Kadlecova2024}; also experimental constraints have been proposed through interferometry \cite{Schellstede2015}.
It is also worth noting that there is a connection between BI-NLED with string theory; in bosonic open string theory
the BI action arises as an effective low-energy action \cite{Fradkin1985}. In \cite{Babaei-Aghbolagh:2022uij} it has shown that the BI non-linear electrodynamic theory is recovered by considering the $T\bar{T}$-deformation of Maxwell theory.

On the other hand, the EH theory was derived from QED principles by W. Heisenberg and H. Euler in 1936 \cite{EH}  from the perspective of QED  \cite{Dunne2021, Sasorov2021, tzenov2020dispersion}. It considers vacuum polarization to one loop and is valid for electromagnetic fields that change slowly compared to the inverse of the electron mass.  By treating the vacuum as a medium, EH effective action predicts rates of nonlinear light interaction processes.
The Lagrangian for EH NLED is 
\begin{equation}
L_{EH}=-\frac{F}{4}+\frac{\alpha^2}{90 m_e^4}\left(F^2+\frac{7}{4}G^2\right),
\label{LEH}
\end{equation}
where $\alpha$ is the fine structure constant and $m_e$ the electron mass.

An interesting aspect of NLED are the symmetries that are preserved.
The BI NLED is not conformally invariant while the EH NLED is neither conformal invariant nor dual invariant. The question then arose if a NLED fulfilling the same symmetries as Maxwell's existed; i.e. a NLED with conformal and electric-magnetic duality invariances. It indeed exists. Recently, a NLED theory endowed with the two symmetries \cite{Bandos2020} was derived.  It is characterized by a dimensionless parameter $\gamma$ and it reduces to Maxwell theory if $\gamma=0$. This theory, known as Modified Maxwell (ModMax, MM) NLED, has stimulated research in several aspects, from classical solutions \cite{Banerjee2022, Neves2023} to supersymmetric analysis \cite{Escobar2022, Bandos2021p, BandosSusy, Barrientos2022, Bordo2021, Bokulic2021, Pantig2022, Babaei2022}. Like the BI non-linear electrodynamic theory, in \cite{Babaei-Aghbolagh:2022uij} the authors showed that the ModMax is recovered by considering the $T\bar{T}$-deformation of Maxwell theory. The Lagrangian for the ModMax NLED is 
\begin{equation}
L_{MM}=-\frac{F}{4}\cosh \gamma +\frac{\sinh \gamma}{4}\sqrt{F^2+G^2}.
\label{LMM}
\end{equation}

In this work, we discuss the phase velocities of light moving in the three previously described NLEDs; although the parameter of each NLED is different we would like to know which one is the most effective in slowing down light velocity in a magnetic or electric background. Additionally, 
there is not a criterion to constrain the values of the ModMax parameter $\gamma$, apart from being positive, as energy conditions of a MM black hole demand.   We use BI and EH theories to set bounds on $\gamma$ through the comparison of the phase velocities of light propagating in a magnetic (or electric) background.  In the second part of the paper, regarding NLED coupled to gravity, we compare the shadows of the respective black hole solutions; confronting their shadows with the observational evidence from Sagittarius $A^*$ and the M87$^*$ galaxy
allows to delimit the ranges of the nonlinear parameters ($\mu$ for EH, $b$ for BI, and $\gamma$ for ModMax) that would fit with charged astrophysical supermassive black holes.

The paper is organized as follows: in Section \ref{Sec:NLED effective metric} it is presented the effective or optical metric, that is the background metric modified by the nonlinear electromagnetic effects; the null geodesics of the effective metric are the actual trajectories of light propagation; then the phase velocity general expression is derived. In Section \ref{Sec:Comparison of NLEDs}, the effective metric and phase velocity of light propagation in an intense magnetic background are given for each one of the NLEDs under study. Through phase velocities for the BI and EH NLEDs we set bounds on the  ModMax nonlinear parameter $\gamma$.
In Section \ref{Sec:BH Shadows}, we consider the static spherically black hole solutions of the three analyzed NLEDs and calculate the corresponding shadows that are then confronted versus the Sagittarius $A^*$ and the M87 galaxy observed shadows. These tests set bounds on the possibility of a charged supermassive black hole. Finally, the conclusions are given in the last section.


\section{NLED effective metric and phase velocity}
\label{Sec:NLED effective metric}

It is well known that intense EM fields, where the Maxwell theory is no longer valid, can resemble a curved spacetime, in the sense that light trajectories are not straight lines but undergo deflection. Deviations from the straight trajectories in vacuum are described in NLED by the null trajectories of an effective or optical metric. The concept of effective metric in nonlinear electrodynamics has its origin in the works by  G. Boillat \cite{Boillat1970}, Bialynicki-Birula \cite{Bialynicka1970},   Pleba\'nski \cite{Pleban} and more recently by Novello \cite{Novello2000}. In this work, we take the approach developed by Pleba\'nski and later on by Novello and collaborators. 

The equations of the propagation of the field discontinuities in nonlinear electrodynamics by a Lagrangian ${L}(F, G)$  can be derived by analyzing the propagation of linear waves associated with the discontinuity of the field, $f_{\alpha\beta}$,  in the limit of geometrical optics \cite{Novello2000b}.

 Considering a null vector $k_{\mu}=(\omega,\vec{k})$ that is normal to the characteristic surfaces or wavefronts, the effective  metric $g_{\rm eff}^{\mu \nu}$ must be the metric in which the wave vector is null, $k^{\mu}k_{\mu}=0,$
\Ari{
\begin{equation}
\label{BIeffgen}
 g_{\rm eff (a)}^{\mu\nu}k_{\mu}k_{\nu}=0, \quad a=1,2,
\end{equation}
}
\Ari{where the $(a)$ subscript} corresponds to the two metrics that can arise in NLED when birefringence occurs.

\Ari{The equations of the propagation of the field are derived by analyzing the behavior of the field discontinuities in the geometrical optics limit \cite{Novello2000}. Starting from the following pair of coupled equations:
\begin{equation}
\label{BIeffgen1}
   \zeta k^2=\frac{4}{L_{F}}F^{\lambda \nu}F^{\mu}{}_{\lambda}k_{\nu}k_{\mu}(L_{FF}\zeta + L_{FG}\zeta^{*})-\frac{G}{L_{F}}k^2 (L_{FG}\zeta+L_{GG}\zeta^{*}),
\end{equation}
\begin{equation}
\label{BIeffgen2}
    \zeta^{*}k^2=\frac{4}{L_{F}}F^{\lambda\nu}F^{\mu}{}_{\lambda}k_{\nu}k_{\mu}(L_{FG}\zeta+L_{GG}\zeta^{*})-\frac{G}{L_{F}}k^2(L_{FF}\zeta+L_{FG}\zeta^{*})+2\frac{F}{L_{F}}k^2(L_{FG}\zeta+L_{GG}\zeta^{*}),
\end{equation}
where, $\zeta=F^{\alpha\beta}f_{\alpha\beta}$ , $\zeta^{*}=F^{*\alpha\beta}f_{\alpha\beta}$ and  $k^2=g^{\mu \nu}k_{\mu}k_{\nu}$, as presented in \cite{Novello2000}. Considering $k^2\neq0$, the propagation vector $k^{\mu}$ is non-null in the background spacetime and the relationship between $\zeta$ and $\zeta^{*}$ is obtained as
\begin{equation}
 \Omega_{1}\zeta^{* 2}+\Omega_{2}\zeta \zeta^{*}+\Omega_{3}\zeta^2=0,
 \label{Eq:quadratic}
\end{equation}
where the coefficients $\Omega_1$, $\Omega_2$ and $\Omega_3$ are given by
\begin{eqnarray*}
     \Omega_{1} &=& -L_{FG}+2\frac{F}{L_{F}}L_{FG}L_{GG}-\frac{G}{L_{F}}L_{FG}^2+\frac{G}{L_{F}}L_{GG}^2, \\
     \Omega_{2} &=& L_{GG}-L_{FF}+2\frac{F}{L_{F}}L_{FF}L_{GG}+2\frac{F}{L_{F}}L_{FG}^2 -2\frac{G}{L_{F}}L_{FF}L_{FG}+2\frac{G}{L_{F}}L_{FG}L_{GG},\\
     \Omega_{3} &=& L_{FG}+2\frac{F}{L_{F}}L_{FF}L_{FG}-\frac{G}{L_{F}}L_{FF}^2+\frac{G}{L_{F}}L_{FG}^2 .
\end{eqnarray*}
Solving the Eq. (\ref{Eq:quadratic}) for $\zeta^{* }$, $\zeta^*=\Omega_{(a)}\zeta$ it is obtained \cite{Novello2000b}, where
\begin{equation}
\Omega_{(a)}=\frac{-\Omega_{2}\pm \sqrt{\Omega_{2}^2-4\Omega_{1}\Omega_{3}}}{2\Omega_{1}}, \qquad a=\mp,
\label{omega_a}
\end{equation}
which leads to two possible paths of propagation that indicates the possibility of birefringence, one  path corresponding to each  polarization mode, then the propagation vector can be expressed as }

\begin{equation}
k_{(a)}^2= - \lambda_{a}F^{\tau \mu}F^{\nu}_{\tau}k_{\mu}k_{\nu},\quad a=1,2,
\label{disprelation}
\end{equation}
where  $\lambda_{a}$ can be interpreted as the  birefringence or refractive index of a material medium and is  given in terms of the NLED Lagrangian and its derivatives by,

\begin{equation}
\lambda_{a}=-4\frac{L_{FF}+\Omega_{(a)}L_{FG}}{L_{F}+G(L_{FG}+\Omega_{(a)} L_{GG})},\quad a=1,2,
\label{MMrefindices}
\end{equation}
where  ${L}_{X}=\frac{d L}{dX}$, $X={F, G}$, and $\Omega_{(a)}$ in Eq. (\ref{MMrefindices}), depends on the derivatives of the Lagrangian with respect to the invariants \cite{Novello2000b}.  The expression for the effective metric can be identified from the dispersion relation, Eq. (\ref{BIeffgen}) with Eq. (\ref{disprelation}) is, 

\begin{equation}
\left\{ g^{\mu\nu} + \lambda_{a}t^{\mu\nu} \right\} k_{\mu}k_{\nu}= 
 g_{{\rm eff}  (a)}^{\mu\nu}k_{\mu}k_{\nu}=0,\quad a=1,2,
\label{MMnullgeod2}
\end{equation}
where $t^{\mu \nu}=F^{\mu \lambda }F_{\lambda}{}^{\nu}$, is defined by the particular (intensive) fields $F_{\mu \nu}$ of the system, and $g^{\mu\nu}$  is the background metric that can be the Minkowski metric or a general curved spacetime, for instance, a black hole or a cosmological spacetime. 
Then  the effective metrics for general nonlinear electrodynamics with Lagrangian $L(F,G)$ are given by,
\begin{equation}
  g_{{\rm eff} (a)}^{\mu\nu} =  g^{\mu\nu}+ \lambda_{a} t^{\mu\nu}, \quad a=1,2.
\label{MMeff_metric}
\end{equation}

If the Lagrangian is such that ${L}_{FG}=0$, \Ari{the index $\lambda_{a}$ reads as 
\begin{equation}
\lambda_{a}=-4\frac{L_{FF}}{L_{F}+G\Omega_{(a)} L_{GG}},\quad a=1,2,
\end{equation}
with the expressions for $\Omega_{(a)}$ simplified accordingly. }
Additionally, when we turn off either the electric or the magnetic field, i.e., $\vec{E}=0$ or $\vec{B}=0$, then $G=4\vec{E}\cdot\vec{B}=0$, and the effective metrics can be obtained from Eqs. (\ref{BIeffgen1}) and (\ref{BIeffgen2}), and the effective metrics become




\begin{eqnarray}
g_{{\rm eff} (1)}^{\mu\nu} &=&  L_{F} g^{\mu\nu}, \\
g_{{\rm eff} (2)}^{\mu\nu} &=&  L_{F}g^{\mu\nu}-4L_{FF}t^{\mu\nu},
\end{eqnarray}
one of the effective metrics, $g_{{\rm eff} (1)}$, is conformal to the background metric; in this case, one of the optical paths is the null geodesic of the background metric and the other one is the null geodesic of the effective metric $g_{{\rm eff} (2)}$. 

In this work, we analyze three different NLEDs: BI, EH and ModMax, and here we present their respective effective metrics, as listed below. 

Note that the BI effective metric does not present birefringence, then there is only one,
\begin{equation}
g_{\rm eff, BI}^{\mu\nu}=\left(b^2+\frac{F}{2} \right) g^{\mu\nu}+F^{\mu}{}_{\lambda}F^{\lambda\nu}.
\label{BI_eff}
\end{equation}

On the other hand, the EH NLED has two effective metrics, given by
\begin{equation}
g_{\rm eff(1), EH}^{\mu\nu} = - \frac{1}{4} \left( 1 + \frac{\mu F \left( \frac{5}{2} F^2 - \frac{7}{4} G^2\right)}{ \left(F^2 + \frac{7}{4} G^2\right)^{3/2}} \right) g^{\mu\nu}
 - \frac{7}{4} \frac{\mu F^2}{ \left(F^2 + \frac{7}{4} G^2\right)^{3/2}} F^{\mu}{}_{\lambda}F^{\lambda\nu},
 \label{EH_eff1}
\end{equation}
and

\begin{equation}
g_{\rm eff(2), EH}^{\mu\nu}= - \frac{1}{4} \left( 1 - \frac{\mu F }{ \left(F^2 + \frac{7}{4} G^2\right)^{1/2}} \right) g^{\mu\nu}
 - \frac{7}{4} \frac{\mu G^2}{ \left(F^2 + \frac{7}{4} G^2\right)^{3/2}} F^{\mu}{}_{\lambda}F^{\lambda\nu},
 \label{EH_eff2}
\end{equation}
where  we  denote $\mu = 2 \alpha^2/(45 m_e^4)$.

Finally, ModMax NLED does present birefringence but one of the two effective metrics is conformal to the background metric; such that the effective metrics are given by

\begin{equation}
g_{\rm eff(1), MM}^{\mu\nu}=g^{\mu\nu}+\frac{4\tanh \gamma}{\sqrt{F^2+G^2}+F\tanh \gamma}t^{\mu\nu},
\label{MM_eff1}
\end{equation}

\begin{equation}
    g_{\rm eff(2), MM}^{\mu\nu}=g^{\mu\nu}.
    \label{MM_eff2}
\end{equation}
Note that the electromagnetic tensor $F_{\mu \nu}$ and its invariants that appear in the previous equations correspond to the intense electromagnetic background field that the light wave traverses with the phase velocity derived from the corresponding $g_{\rm eff}^{\mu\nu}$.
In the Maxwell case  $L=-F/4,\quad L_{F}=-1/4, \quad L_{FF}= 0 $ and $L_{G}= 0$, then both effective metrics become conformal to the background metric, $g_{\rm eff}^{(1) \mu\nu} = g_{\rm eff}^{(2) \mu\nu}= g^{ \mu\nu}$, and the null geodesics coincide with the light trajectories in the background spacetime.

See \cite{Obukov2002} for a study on the Fresnel equation in nonlinear electrodynamics and \cite{Goulart2009} for a classification of the effective metrics.

With the nonlinear Lagrangians, one can also obtain the Plebanski dual variable \cite{Pleban}
\begin{equation}
    P_{\mu\nu}= -4 L_{F}F_{\mu\nu}-4L_{G}F^{*}_{\mu\nu}.
\end{equation}
For the NLEDs in consideration 
\begin{equation}
P_{BI,\mu\nu}= \frac{1}{\sqrt{1+\frac{F}{2b^2}-\frac{G^2}{16b^4}}} \left(F_{\mu\nu}-\frac{G}{4b^2}F^{*}_{\mu\nu} \right),
\end{equation}

\begin{equation}
    P_{EH,\mu\nu}= F_{\mu\nu} -2\mu \left(F F_{\mu\nu}+\frac{7 G}{4} F^{*}_{\mu\nu} \right),
\end{equation}

\begin{equation}
    P_{MM,\mu\nu}=\left(\cosh \gamma - \frac{F \sinh \gamma}{\sqrt{F^2 +G^2}}F_{\mu\nu} \right) - \frac{G \sinh \gamma}{\sqrt{F^2 +G^2}}F^{*}_{\mu\nu}.
\end{equation}


\subsection{Phase velocity of light propagation in the Effective Metric}

Here we explain how the phase velocity of light can be calculated
from the dispersion relation.  Let us consider a  null vector $k_{\mu}=\{ \omega, \vec{k} \}$. If $k_{i}$ is the light propagation  direction, from the dispersion relation,  Eq. (\ref{BIeffgen}), 

\begin{equation}
g_{{\rm eff}(a)}^{tt}\omega^{2}+2g_{{\rm eff}(a)}^{it}\omega k_{i}+g_{{\rm eff}(a)}^{ij} k_i k_j=0,
\label{MMdr1}    
\end{equation}
where $i,j$ subscripts (or superscripts) denote the spatial coordinates and $t$ denotes the time coordinate. Defining the normalized wave vector $\tilde{k}_{i}=k_{i}/|\vec{k}|$, Eq. (\ref{MMdr1}) can be written  as 

\begin{equation}
    g_{{\rm eff}(a)}^{tt}\frac{\omega^{2}}{|\vec{k}|^2}+2g_{{\rm eff}(a)}^{it}\frac{\omega}{|\vec{k}|} \tilde{k}_{i}+g_{{\rm eff}(a)}^{ij} \tilde{k}_{i}\tilde{k}_{j} =0.
\end{equation}

Then the light's phase velocity,  $v=\omega/|\vec{k}|$, for propagation in the direction $\tilde{k}_{i}$,  $v^{i}$ is given by 
\begin{equation}
   \left(v^{i} \right)_{a}= \frac{\omega}{|\vec{k}|}\tilde{k}_{i}=\frac{g_{{\rm eff}(a)}^{it}\tilde{k}_{i}}{g_{{\rm eff}(a)}^{tt}}\pm\sqrt{\left(\frac{g_{{\rm eff}(a)}^{it}\tilde{k}_{i}}{g_{{\rm eff}(a)}^{tt}}\right)^2-\frac{g_{{\rm eff}(a)}^{ij}\tilde{k}_{i}\tilde{k}_{j}}{g_{{\rm eff}(a)}^{tt}}}, \qquad a=1,2.
    \label{MMphase_velocity}
\end{equation}

The two possible effective metrics $g_{{\rm eff}(a)}^{\mu\nu}$, $a=1,2$, render two
dispersion relations that correspond to two modes of polarization, this is known as the birefringence effect \cite{Novello2000}. Note that these velocities differ from the phase velocities of a massless test particle that follows the null geodesics of the background metric $g_{\mu\nu}$ and  is given by analogous expressions making $g_{\rm eff(a)}^{\mu\nu} \mapsto g^{\mu\nu}$, 

\begin{equation}
    v_{\rm massless} = \pm \sqrt{-\frac{g^{ij}\tilde{k}_{i}\tilde{k}_{j}}{g^{tt}}}.
\end{equation}

\section{Comparative of three nonlinear electrodynamics: Light propagation}
\label{Sec:Comparison of NLEDs}

In this Section, our goal is to study light propagation in the three NLED theories and to evaluate how strong electromagnetic fields impact its phase velocity. We will compare the phase velocities of light propagating in a magnetic background in a Minkowski spacetime, and then in an electric background. 

\subsection{NLED phase velocities in a magnetic background }
We examine the phase velocities of light propagating in the $z$ direction, while the magnetic background field is oriented making an angle with the propagation direction; we consider the magnetic field with one component along the propagation direction and a second component as transversal or perpendicular,  $\vec{B}=  B_z \hat{z} + B_{x} \hat{x} + B_{y}\hat{y} =B_z \hat{z} + B_{\perp} (\hat{x} + \hat{y})$. We calculate the phase velocities for the case of a static observer or as measured in the Laboratory.
The phase velocities corresponding to the three NLEDs are derived from Eq. (\ref{MMphase_velocity}),  considering a Minkowski background,
$g^{\mu \nu}= \eta^{\mu \nu} = {\rm diag} [-1,1,1,1]$; while the uniform magnetic field components are $F^{xy}=-B_{z},\quad F^{xz}=B_{y}, \quad F^{yz}=-B_{x}$, and the invariants, $ F= 2B^2, \quad G=0$.

For the EH NLED, the expressions for the phase velocities are
\begin{equation}
v_{\text{EH1}}^2=1-\frac{14 \mu  B_{\perp}^2}{10 B^2 \mu +1}, \quad v_{\text{EH2}}^2=1-\frac{8 \mu  B_{\perp}^2}{1-4 B^2 \mu }.
\end{equation}
Considering terms up to $\mu$ ( $\alpha^2$), the EH phase velocities are given by,

\begin{equation}\label{EHphvel}
v_{\text{EH1,B}}^2=1- 14 \mu B_{\perp}^2, \quad  v_{\text{EH2,B}}^2=1- 8\mu B_{\perp}^2.  
\end{equation}
And performing an average over polarization and direction,
\begin{equation}\label{EHphvel_prom}
v_{\text{EH,B}}^2=1- \frac{11 \alpha}{270 \pi} \left( \frac{B_{\perp}}{B_{cr}} \right)^2,  
\end{equation}
where we substituted $\mu$ in terms of the critical field  $\mu= \alpha/(90 \pi B_{cr}^2)$; that gives a light velocity decreasing respect the one in vacuum $c=1$ of
\begin{equation}\label{EHphvel_prom_n}
v_{\text{EH,B}}^2=1- (9.5 \times 10^{-5}) \left( \frac{B_{\perp}}{B_{cr}} \right)^2. 
\end{equation}

For  the BI NLED, there is no birefringence, and the phase velocity is 
\begin{equation}
    v_{\text{BI}}^2=1-\frac{B_{\perp}^2}{b^2+B^2},
\end{equation}
with $b$ being the maximum attainable electric or magnetic field.
Up to first order in $1/b^2$, neglecting terms $1/b^4$ and higher, 

\begin{equation}\label{BIphv}
    v_{\text{BI,B}}^2=1-\frac{B_{\perp}^2}{b^2}.
\end{equation}
Such that light velocity is reduced the most, for fields with a component perpendicular to the light propagation ($B_{\perp}$) of intensities of the order of $b= 10^{11}$ Tesla.
For the ModMax NLED, there is birefringence and one of the light trajectories is 
the one in vaccum, with phase velocity $v^2=c^2=1$. While the other one is
\begin{equation}
v_{\text{MM}}^2=1-\frac{2 B_{\perp}^2}{B^2} \left( \frac{\tanh(\gamma)}{1 + \tanh(\gamma )} \right).
\label{vmmB}
\end{equation}

\begin{figure}[t]
    \centering
\subfigure{\includegraphics[width=0.7\textwidth]{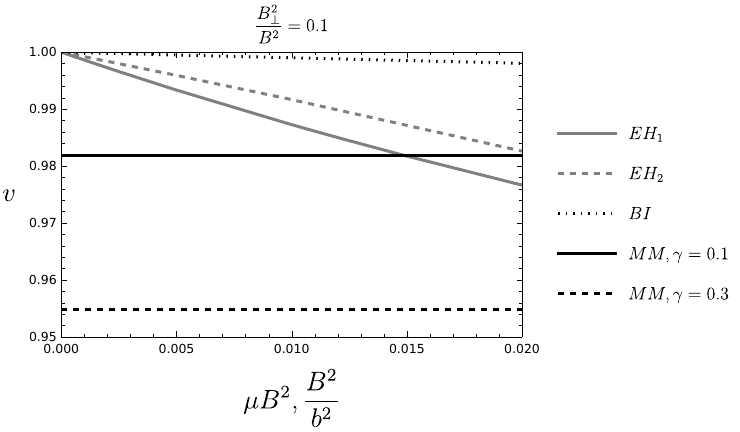}}
    \subfigure{\includegraphics[width=0.67\textwidth]{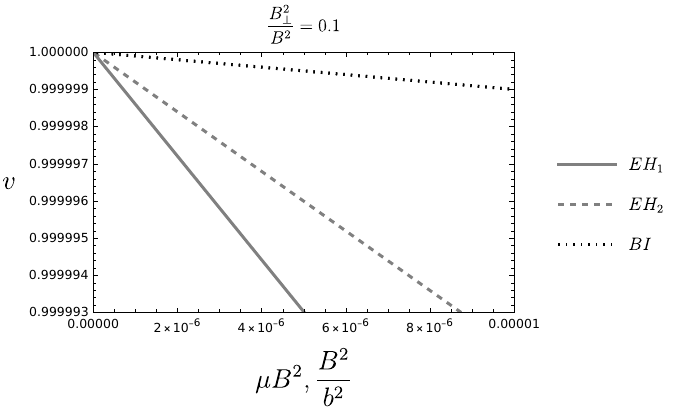}}
    \caption{\footnotesize The phase velocities in the $z-$direction for three NLEDs are plotted. The gray lines correspond to a wave's phase velocity in the presence of an EH magnetic field. The black dotted line corresponds to the phase velocity of a wave in the presence of a BI background field. The black solid and dashed lines 
    corresponds to the value for the light velocity in vacuum in ModMax NLED. The lower panel is the zoom for the differences between  EH and BI, which are qualitatively similar. In the plot $B_{\perp}^2/B^2=0.1$.}
    \label{fig:compEHBIMM1}
\end{figure}

\begin{figure}[t]
\centering
\includegraphics[width=0.8\linewidth]{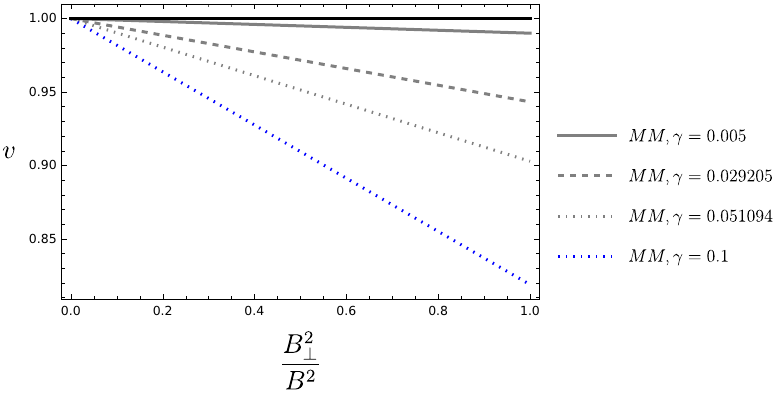}
\caption{\footnotesize The phase velocity for the ModMax NLED is plotted for values of $\gamma$ in the interval established by the BI NLED.  $\gamma=0.005$ comes out considering that the magnetic field is one-tenth of the BI magnetic field, $B/b = 10^{-1}$ and $\gamma=0.1$ corresponds to  a field $B/b = 0.446471$ }
\label{fig:MMbounds}
\end{figure}

The phase velocities as functions of the corresponding adimensional parameters: $\mu B^2$ for EH; $B^2/b^2$ for BI are illustrated in Fig. \ref{fig:compEHBIMM1}. For the ModMax NLED the phase velocity depends on $B_{\perp}/B$; in Fig. \ref{fig:compEHBIMM1}  is shown how the phase velocity decreases in ModMax case for $B_{\perp}/B =0.1$.    The differences in the decreasing of the phase velocities is qualitatively the same for BI and EH, while the Mod Max case is illustrated as a function of $B_{\perp}/B$ in Fig. \ref{fig:MMbounds}. 
\subsection{NLED phase velocities in an electric background }
The phase velocities when considering an electric background field in a Minkowski spacetime are presented. The setting is analogous to the one for the magnetic background: light propagating in the $\hat{z}$ direction and the electric background of the form $\vec{E}= E_z \hat{z} + E_{\perp} (\hat{x} + \hat{y})$. 

The two possible phase velocities for the EH NLED are given by 
\begin{equation}
v_{\text{EH1,E}}^2=1-\frac{14 E_{\perp}^2 \mu }{4 E^2 \mu +1}, \quad v_{\text{EH2,E}}^2=1-\frac{8 E_{\perp}^2 \mu }{12 E^2 \mu +1}.
\end{equation}
For the BI NLED, since there is no birefringence, the phase velocity reads
\begin{equation}
    v_{\text{BI,E}}^2=1-\frac{E_{\perp}^2}{b^2}.
\end{equation}
While  phase velocity in a ModMax NLED is 
\begin{equation}
v_{\text{MM,E}}^2=1-\frac{2 E_{\perp}^2}{E^2} \left( \frac{\tanh(\gamma)}{1 + \tanh(\gamma )} \right).
\label{vmmE}
\end{equation}
The behavior at the order $\mu E^2, \frac{E^2}{b^2} \sim 10^{-2}$ is very similar to the one for the magnetic background field. Moreover, the phase velocities for the wave in the presence of ModMax magnetic and electric fields in Eqs. (\ref{vmmB}) and (\ref{vmmE}) have the same dependence on the magnitude of the fields, a consequence of the duality invariance required by ModMax theory.  

\subsection{Bounds on the ModMax parameter $\gamma$ from BI and EH}

The phase velocity of light propagation in a strong magnetic field background obtained from the ModMax theory is given by,
\begin{equation}
v_{\text{MM}}^2=1-\frac{2 {B_{\perp}}^2}{B^2 } \left( \frac{ \tanh (\gamma )}{1 + \tanh (\gamma )} \right).
    \label{vmmB1}
\end{equation}
As a consequence that there are no bounds for the allowed values of $\gamma$, and that for large values of $\gamma$, $\tanh(\gamma) \mapsto 1$, in the expression for the phase velocity, the factor with $\gamma$ tends to $1/2$ and then $v_{\text{MM}}$ can be made arbitrarily small, even for small values of the magnetic field if the component that is perpendicular to the wave propagation is of the order of the total magnetic field, $B_{\perp} \approx B$. 
\Ari{Since EH describes nonlinear electromagnetic effects like vacuum polarization that induces in turn the effect of photon splitting, the phase velocity of light propagation derived from these effects compared with the one from the ModMax theory can give insight of the ranges of $\gamma$. For this reason, we search for bounds on $\gamma$ from the phase velocities derived from EH, that is a theory that has been experimentally confirmed; while the comparison with BI, a theory that eventually is going to be confirmed as well \cite{Schellstede2015},  can also give a bound for $\gamma$.}
\subsubsection{Bounds on the NLED ModMax parameter  from BI}

Comparing the phase velocity for light propagation that arises in the BI theory  in a magnetic background, Eq. (\ref{BIphv}) with the phase velocity  from the ModMax theory, Eq. (\ref{vmmB1}),  it is obtained for the parameter $\gamma$

\begin{equation}
\tanh{\gamma} = \frac{B^2}{2 b^2- B^2} \approx    \frac{B^2}{2 b^2},
\end{equation}
if we consider a magnetic field background of the order of the maximum attainable magnetic field $b$, i.e. $B/b \approx 1$, we obtain the bound 
$\gamma= 0.549306$. While if  $B$ is a tenth of $b$, $B/b \approx 0.1$ then the bound on the ModMax parameter is $\gamma= 0.005$. Smaller magnetic fields give smaller bounds for $\gamma$.
Then BI constrains $\gamma$ to the interval $0 < \gamma \le 0.549306$.


\subsubsection{Bounds on the NLED ModMax parameter  from EH}

The two phase velocities for light propagation in a strong magnetic background that arise in the EH theory, Eqs. (\ref{EHphvel}), are compared with the MM phase velocity.  From the first one, $v_{\text{EH1,B}}$  compared with the phase velocity for the ModMax theory, Eq. (\ref{vmmB1}), we obtain for the parameter $\gamma$ the constraint

\begin{equation}
\tanh{\gamma} = \frac{7 \alpha}{90 \pi} \left(\frac{B}{B_{cr}} \right)^2 = 1.8 \times 10^{-4} \left(\frac{B}{B_{cr}} \right)^2, 
\end{equation}
Such that for fields near the critical one, $B \approx B_{cr}$
the nonlinear MM parameter is $\gamma = 1.8 \times 10^{-4}. $

Following the same procedure for the second velocity  $v_{\text{EH2,B}}$ in Eqs.  (\ref{EHphvel}) we obtain a second bound of $\gamma = 10^{-4},$
for fields $B \approx B_{cr} = 10^{9}$ Tesla.
Such that the EH theory sets 
$\gamma$ in the interval $0.0001 < \gamma < 0.00018$.

Therefore the  EH theory delimits the nonlinear ModMax parameter to the interval $0.0001 < \gamma < 0.00018$, while BI places  $\gamma$ in a wider interval, $0 \le \gamma \le 0.549306$. We should take into account that EH and BI apply to different nonlinear electromagnetic effects that arise at different orders of magnitude of the background fields
( $B_{cr} \approx 10^{9}$ Tesla for EH and $b \approx 10^{11}$ Tesla for BI), 
but both serve to the purpose of delimiting the value of $\gamma$, i.e. the effect of the ModMax electrodynamics in light propagation 
should resemble in a certain way the ones of the BI or EH electrodynamics. The BI constraint $0 < \gamma < 0.549306$  includes the constraints derived from EH,  therefore the greater field intensity allows a larger $\gamma$ in the ModMax theory.

\section{Comparative of three nonlinear electrodynamics: Black Hole Shadows}
\label{Sec:BH Shadows}

It is well known that the coupling of Nonlinear Electrodynamics (NLED) with Einstein's gravity can yield black hole solutions. In this section, we focus on the static black holes corresponding to the three NLEDs discussed above and we compare the metric functions for these BHs with the Reissner-Nordstrom one, which is the linear limit; the comparison also includes the Schwarzschild BH as a reference. Additionally, we compare the resulting shadow radii with the observed shadows of Sagittarius $A^*$ and M87$^*$ black holes \cite{Vagnozzi2023,Chakhchi2024,EHT_M87}. In this regard, see \cite{Allahyari:2019jqz}, where one of the first comparisons against observations of M87$^*$ coming from EHT was made to study the shadows of non-linear electrodynamics BHs. 
\subsection{ NLED BH Metric functions}
We consider the corresponding BH metrics as the background metrics. It is interesting to compare the metric functions for the BHs of the three NLEDs in consideration with the one of  Reissner-Nordstrom, the static spherically symmetric solution of the coupled Einstein-Maxwell equations. For the static spherically symmetric line element, that is the background metric $g_{\mu \nu}$,
\begin{equation}
ds^2= - f(r) dt^2 + \frac{dr^2}{f(r)} + r^2 d \Omega,   
\end{equation}

the Reissner-Nordstrom (RN) metric function is given by
\begin{equation}
f_{RN}(r) = 1-\frac{2 M}{r}+\frac{Q^2}{r^2},
\end{equation}
characterized by BH mass, $M$, charge, $Q$, and electromagnetic field
\begin{equation}
F_{rt}= \frac{Q}{r^2}=P_{rt}.    
\end{equation}
We consider as a reference the Schwarzschild (Schw) BH, that is chargeless and with metric function $f(r)$ given by,
\begin{equation}
f_{Schw.}(r) = 1-\frac{2M}{r}.    
\end{equation}
As the nonlinearly charged BH we analyze the Euler-Heisenberg (EH) BH; the dyonic solution, characterized by the BH mass, M, and the electric and magnetic charges, $Q_e$ and $Q_m$, was presented in \cite{Ruffini2013} and the metric function $f(r)$ and electromagnetic field components are, respectively,
\begin{eqnarray}
f_{EH}(r) &=& 1-\frac{2 M}{r}+\frac{Q^2}{r^2} - \frac{\mu  Q^4}{20 r^6},\\
E(r) &=& \frac{Q_e}{r^2} \left[ 1- \frac{\mu}{2 r^4} \left(Q_e^2+ \frac{5}{2} Q_m^2 \right)\right], \\
B(r) &=& \frac{Q_m}{r^2} \left[ 1 + \frac{\mu}{2 r^4} \left(Q_m^2+ \frac{5}{2} Q_e^2 \right)\right],
\end{eqnarray}
recalling that $\mu = 2 \alpha^2/(45 m_e^4)$.\\
The second case analyzed is the Born-Infeld (BI) BH, with
metric function $f(r)$ 
\begin{equation}
     f_{BI}(r) = 1-\frac{2 M}{r}+\frac{2}{3} b^2 r^2 \left(1-\sqrt{\frac{Q^2}{b^2 r^4}+1}\right)+\frac{2 Q^2 }{3r}\sqrt{\frac{b}{Q}} F\left[{\rm ArcCos} \left(\frac{\frac{b r^2}{Q}-1}{\frac{b r^2}{Q}+1}\right),\frac{1}{\sqrt{2}}\right],
\end{equation}
with electromagnetic field given by
\begin{equation}
F_{rt}(r)=\frac{ \sqrt{Q^2}}{\sqrt{r^4+ Q^2/b^2}}, \quad P_{rt}(r)= \frac{Q^2}{r^2};
\end{equation}
where $Q^2=Q_e^2+Q_m^2$; the geodesic structure of this solution was presented in \cite{breton2002geodesic}.
Finally, the Mod Max BH was derived in \cite{Maceda2020}, with the line element
\begin{equation}
f_{MM}(r) = 1-\frac{2 M}{r}+\frac{e^{- \gamma} Q^2}{r^2},
\end{equation}
where $Q^2=Q_e^2+Q_m^2$; and it is characterized by the electromagnetic field
\begin{equation}
F_{\mu \nu} (r)= \delta_{[\mu}^{t} \delta_{\nu]}^{r} \frac{Q_e e^{- \gamma}}{r^2}- Q_m \sin \theta \delta_{[\mu}^{\theta} \delta_{\nu]}^{\phi}.     
\end{equation}

\Ari{The equation that defines the horizons, $f(r)=0$, is complicated for both the BI and EH cases, requiring numerical methods to solve. However, in the case of the EH black hole, Descartes' rule of signs can be applied to show that the equation $f(r)$ has three positive roots \cite{Allahyari:2019jqz,Magos2020}. Since there are three sign changes, this confirm the existence of three positive solutions for $f(r)=0$, }
\Ari{
\begin{equation}
    f_{EH}(r) = 1 -\frac{2 M}{r}+\frac{Q^2}{r^2} - \frac{\mu  Q^4}{20 r^6}  = 0.
\end{equation}
   }
\textcolor{black}{There exist then one external and two inner horizons; although not the most common situation, in the literature there are several cases of BH with more than two horizons.
For BH with a de Sitter asymptotics it arises an apparent horizon known as the cosmological horizon.   In the EH case the third horizon dissapears when $\mu$ vanishes, and the RN case with two horizons is recovered \cite{Magos2020}. Also  when an additional scalar field is introduced or in nonlinear electromagnetism \cite{Gao:2021kvr}
the BH may have more than two horizons. }

\begin{figure}[t]
\centering
\includegraphics[width=0.8\textwidth]{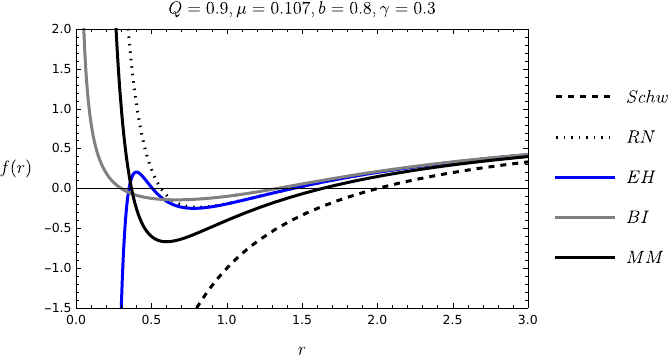}
\caption{\footnotesize The metric functions $f(r)$ for the black holes considered are shown. The radii of the \Ari{Killing horizons} are the intersections of the corresponding metric functions with the $r-$axis. }
\label{fig:discfr}
\end{figure}

In Fig. \ref{fig:discfr}, we plot the metric functions for the Schwarzschild, RN, EH, BI, and ModMax black holes, for a fixed  BH charge $Q=0.9$. In the BI case for $b>0.7$, $f_{BI}(r)$ behaves similarly near the origin than the one for RN, while if $b<0.7$, the metric function $f_{BI}(r)$ resembles the  Schwarzschild behavior near $r=0$ \cite{breton2002geodesic}.

The magnetic field does not contribute to the pair production rate, so that the process of electron-positron  pair production does not occur for a strong magnetic field. However if the electric field exceeds the critical value for the Schwinger pair creation the BH can be destabilized; in \cite{Ruffini:1999kq} the pair creation arising when the electric field exceeds the critical value is linked to the origin of energetic gamma ray bursts. It is worth to stress that the impact of the NLED effects strongly depends on the BH mass: the more massive the BH is, the bigger is the charge needed to trigger the NLED effects; for instance, the EH effects are inversely proportional to the BH mass.   Moreover, a limit  on the BH mass  $M_{\rm max} = 7.2 \times 10^{6} M_{\odot}$, was established by requiring that the pair production process would last less than the age of the Universe. For masses much smaller than $M_{\rm max}$ the pair production process  can drastically modify the electromagnetic structure of the BH.
As we show in the next subsection,  the comparison of the NLED shadows with observational evidence of the shadows of Sagittarius $A^{\ast}$ and $M87^{\ast}$ works well for $Q/M \le 0.35$,
while for the considered BH masses the range of $Q/M \le 0.35$ is in agreement with the bounds derived from NLED effects.

Based on the irreducible mass, $M_{\rm ir}$, that is the BH mass that cannot be reduced by any process of energy extraction,  an upper bound  on the BH charge can be derived by considering a loss of charge by pair production \cite{Christodoulou:1971pcn}; the bound is given by
\begin{equation}
\frac{Q^4}{16 M_{\rm ir}^4} \le 1;
\label{IneqQM}
\end{equation}
once the equality is satisfied, no more charge can be added to the BH without increasing the irreducible mass.    For a static spherically symmetric (SSS) BH $M_{\rm ir}$ is given by
\begin{equation}
M_{\rm ir}^{SSS} = \frac{r_{+}^2}{2}, 
\end{equation}
where $r_{+}$ is the radius of the BH horizon. 
In general the horizon radius of a NLED BH is larger than the RN horizon, 
$r^{NLED}_{+} > r^{RN}_{+}$, then the irreducible mass of a NLED BH is larger than the RN one,
\begin{equation}
M^{NLED}_{\rm ir} = \frac{r^{NLED}_{+}}{2} > \frac{r^{RN}_{+}}{2} = M^{RN}_{\rm ir} \approx \frac{M}{2},
\end{equation}
so it is expected that for a NLED BH Eq. (\ref{IneqQM}) be more stringent than for the RN BH.
For  the ranges we deal with, $Q/M \le 0.35$, there is agreement with the bounds from the irreducible BH mass,
\begin{equation}
\frac{Q^4}{16 M_{\rm ir}^4} < \frac{Q^4}{ M^4} \le (0.35)^4 = 1.5 \times 10^{-2} < 1, 
\end{equation}

Concerning the extraction of electromagnetic energy from a BH, 
Damour and Ruffini \cite{TDamour1975}  showed that a large fraction of the energy of an electromagnetic black hole (EMBH) can be extracted by pair creation. However this process can only work for EMBH with $M_{\rm ir} \le 10^{6} M_{\odot}$. In this respect there are two conceptually different processes, depending on whether the maximal electric field strength is smaller or greater than the electric critical field $E_{cr}$ \cite{Ruffini:2009hg}.
The maximal strength of the electric field is reached at
 the horizon of the BH,  $r_{+}$, 
\begin{equation}
E_{+}(r)= \frac{Q_e}{r_{+}^2}.
\end{equation}
For $E_{+} < E_{cr}$ the leading extraction process consists of a sequence of discrete elementary decay processes of a particle into two oppositely charged particles; regarding the charge-mass ratio of the BH it should be that
\begin{eqnarray}
Q/M & \le &  10^{-6} \frac{M}{M_{\odot}}, \quad {\rm if} \quad \frac{M}{M_{\odot}} \le 10^{6}, \nonumber\\
Q/M & \le &  1, \quad {\rm if} \quad \frac{M}{M_{\odot}} > 10^{6}. \nonumber
\end{eqnarray}
While for $E_{+} > E_{cr}$ the leading extraction mechanism is collective and based on the generation of the optically thick electron-positron plasma by the vacuum polarization. It is shown that in this case
\begin{equation}
2 \times 10^{-6} \frac{M}{M_{\odot}} < \frac{Q}{M} < 1,
\end{equation}
this occurs only for a BH with mass smaller than $5 \times 10^{5}{M_{\odot}}$.
Therefore,  the previous bounds are fulfilled in our case since we kept $Q/M < 1$, even if the NLED case allows $Q>M$ due to the NLED charge screening.
\subsection{The shadow of the NLE black holes}
Considering the equations of motion of a test particle in the external field of a static BH, the radial motion is governed by
\begin{equation}
   \dot{r}^2 
   + \frac{L^2}{g_{rr}} \left( \frac{g_{\phi\phi}}{r^4}
   +\frac{g_{tt}}{f^2 (r)} \frac{1}{b^2} 
   - \frac{\delta}{L^2} \right)
   = 0,
   \label{geodesic}
\end{equation}
where $\delta= 1,0,-1$ for space-like, null, or time-like geodesics, respectively; the test particle angular momentum is $L$,  and  the impact parameter $b= L/E$, where  $E$ is the particle energy.
In NLED photons travel through null geodesics of the effective metric, then  in Eq. (\ref{geodesic}) $\delta=0$ and $g \mapsto g^{{\rm eff}(a)}$, such that the radial motion of the light ray obeys,
\begin{equation}
   \dot{r}^2 
   + \frac{L^2}{g^{{\rm eff}(a)}_{rr}} \left( \frac{g^{{\rm eff}(a)}_{\phi\phi}}{r^4}
   +\frac{g^{{\rm eff}(a)}_{tt}}{f^2 (r)} \frac{1}{b^2} \right)
   = 0, \quad a=1,2,
   \label{MMgeodesic}
\end{equation}
where $a=1,2$ denote the two effective metrics, given by  Eq. (\ref{BI_eff}) for BI, Eqs. (\ref{EH_eff1})- (\ref{EH_eff2}) for EH, and Eqs.   (\ref{MM_eff1})- (\ref{MM_eff2}) for ModMax. The electromagnetic tensor components  $F_{\mu \nu}$ for the construction of the effective metric  correspond to the ones of  a charged static BH, that in case of both charges, electric $Q_e$ and magnetic $Q_m$  are given by 

\begin{equation}
F^{tr} = \frac{Q_e}{r^2 {L}_{F}}, \quad F_{\theta \phi} = Q_m \sin{\theta},    
\end{equation}
we shall consider only electric charged BHs, $Q_e=Q$.  

From 
$\dot{r}^2 + V^{(a)}_{\rm eff} =0,$ we identify the effective potential as 
\begin{equation}
    V^{(a)}_{\rm eff}=\frac{L^2}{g^{{\rm eff}(a)}_{rr}} \left( \frac{g^{{\rm eff}(a)}_{\phi\phi}}{r^4}
   +\frac{g^{{\rm eff}(a)}_{tt}}{f^2 (r)} \frac{1}{b_{a}^2} \right), \quad a=1,2.
   \label{MMeffpotential}
\end{equation}
The radius of the  circular orbits $r_{c}$ 
corresponds to an extreme of the effective potential,
\begin{equation}
V^{(a)}_{\rm eff} |_{r_c} =0, \quad  \frac{dV^{(a)}_{\rm eff}}{dr} |_{r_c}  =0,\quad a=1,2.
\label{MMcirc_orb_conds}
\end{equation}

The critical impact parameters can be obtained in terms of the effective metrics as 
\begin{equation}
(b^2_{c})_{a}= \left( -\frac{r_{c}^4 g_{tt}^{{\rm eff} (a)}}{f^2 (r_{c})g_{\phi\phi}^{{\rm eff} (a)}}\right)_{r_{c}},\quad a=1,2.
\end{equation}
The NLED theories that present birefringence have two values for the critical impact parameter, corresponding to the two effective metrics;  for the NLED-BHs of our study, EH, BI, and ModMax, respectively, are

\begin{equation}
    b_{c,EH1}^2=\frac{r_{c}^2}{f_{EH}(r_{c})} \left( \frac{1-10\mu \frac{Q^2}{r_{c}^4}}{1+4\mu \frac{Q^2}{r_{c}^4}} \right), \quad b_{c,EH2}^2=\frac{r_{c}^2}{f_{EH}(r_{c})} \left( \frac{1+4\mu \frac{Q^2}{r_{c}^4}}{1+12\mu \frac{Q^2}{r_{c}^4}} \right),
\end{equation}

\begin{equation}
    b_{c,BI}^2=\frac{r_{c}^2}{f_{BI}(r_{c})} \left( 1-\frac{Q^2}{b^2 r_{c}^4} \right),
\end{equation}

\begin{equation}
    b_{c,MM1}^2= \frac{e^{-2\gamma}r_{c}^2}{f_{MM}(r_{c})}, \quad b_{c,MM2}^2= \frac{r_{c}^2}{f_{MM}(r_{c})}. 
\end{equation}

We calculate the radii of the shadows for the three NLED-BHs and compare them with the RN BH. The shadow can be determined by considering a light ray from the observer position $r_{o}$ to the unstable circular orbit (UCO) making an angle $\psi_{sh}$ with respect to the radial direction. The expression for the deviation angle associated with the shadow is \cite{Perlick2015}
\begin{equation}
\sin^2\left(\psi_{sh}\right)_{a}=-b^2_{ca}\left(\frac{g_{\phi \phi}^{{\rm eff} (a)}}{g_{tt}^{{\rm eff} (a)}}\frac{f^2 (r_{o})}{r_{o}^4} \right)_{r_{o}}, \quad a=1,2.
\end{equation}
where  $b_c$ is the impact parameter of the UCO with radius $r_c$.

The NLED theories that present birefringence have two values for the shadow radius  $r_{sh}$, corresponding to the two effective metrics. The radius of the BH shadow is approximated in terms of the observer position $r_{o}$ and the angle $\psi_{sh}$ as \cite{Amaro2022,Amaro2020,Perlick2018}
\begin{equation}
r_{sh}=r_{o}\tan \psi_{sh} \approx r_{o}\sin \psi_{sh},
\end{equation}
then, we obtain that for the EH BH there are two radii, then two shadows, 
\begin{equation}
r_{sh,EH1}^2=r_{c}^2 \left(\frac{f_{EH}(r_{o})}{f_{EH}(r_{c})}\right) \left( \frac{r_{c}^4 -10 Q^2 \mu }{r_{c}^4 +4Q^2 \mu}\right) \left( \frac{r_{o}^4 +4 Q^2 \mu}{r_{o}^4 -10Q^2 \mu} \right),
\end{equation}

\begin{equation}
r_{sh,EH2}^2=r_{c}^2 \left(\frac{f_{EH}(r_{o})}{f_{EH}(r_{c})}\right) \left(\frac{r_{c}^4 +4 Q^2 \mu }{r_{c}^4 +12Q^2 \mu} \right) \left( \frac{r_{o}^4 +12 Q^2 \mu}{r_{o}^4 +4 Q^2 \mu} \right).
\end{equation}
The static BI BH has one shadow radius, 
\begin{equation}
 r_{sh,BI}^2 = r_{c}^2 \left( \frac{f_{BI}(r_{o})}{f_{BI}(r_{c})} \right) \frac{ \left( 1 - \frac{Q^2}{b^2 r_{c}^4} \right)}{ \left(1- \frac{ Q^2}{ b^2 r_{o}^4} \right)}.
\end{equation}
Although there are two critical impact parameters for the ModMax BH, an observer located at infinity observes only one shadow,
\begin{equation}
r_{sh,MM}^2= r_{c}^2 \left( \frac{f_{MM} (r_{o})}{f_{MM} (r_{c})} \right).
\end{equation}
Considering that the observer is at infinity $r_{o} \mapsto \infty$, then $f(r_o \mapsto \infty)=1$ and at first order in the nonlinear parameters, the radii of the shadows are, for EH BH up to first order in $\mu$,

\begin{equation}
r_{sh,EH1}^2= \left(\frac{r_{c}^2}{f_{EH}(r_{c})} \right) \left(1 - \frac{14 Q^2 \mu}{r_{c}^4} \right), \quad r_{sh,EH2}^2= \left(\frac{r_{c}^2}{f_{EH}(r_{c})} \right) \left(1 - \frac{8Q^2 \mu}{r_{c}^4} \right).
\end{equation}
For the BI BH up to first order in $1/b^2$,
\begin{equation}
r_{sh,BI}^2= \left(\frac{r_{c}^2}{f_{BI}(r_{c})} \right) \left(1 - \frac{Q^2 }{b^2 r_{c}^4} \right).
\end{equation}
While for the ModMax BH the shadow radius is
\begin{equation}
r_{sh,MM}^2= \left(\frac{r_{c}^2}{f_{MM}(r_{c})} \right).
\end{equation}
It is worth noting that $r_c$ is different for each BH; $r_c$ is the radius of the photosphere,  precisely the radius at which is maximum  the effective potential for the null geodesics of the effective metric;
i.e. $r_c$ is calculated from the conditions $V_{\rm eff}^{\prime} =0 $ and  $V_{\rm eff}^{\prime \prime}  < 0 $. For RN the radius of the UCO is given by

\begin{equation}
r_{c,RN}= \frac{3M}{2} \left( 1 + \sqrt{1- \frac{8 Q^2}{9M^2}} \right),
\end{equation}
Moreover, the impact parameter and shadow radius at infinity of the RN BH can be derived from any of the above expressions in the limit of the vanishing nonlinear parameter, and are given by,
\begin{equation}
b_{c,RN}^2=\frac{r_{c}^2}{f_{RN}(r_{c})} =  r_{sh,RN}^2= \frac{3^3 M^2  \left( 1+ \sqrt{1- \frac{8Q^2}{9M^2}} \right)^4}{8 \left( 1 - \frac{2Q^2}{3M^2} +  \sqrt{1- \frac{8Q^2}{9M^2}} \right)},
\end{equation}
while for Schwarzschild it is well known that the radius of the photosphere $r_c = 3 M$, while the shadow radius is $r_{sh,Schw}= 5.196 M$. 

\subsection{Confronting the NLED BHs shadow with observations}

In this section, we test the NLED BH shadows with the observations of the shadow reported by the EHT for the supermassive BHs in the center of the Milky Way (Sgr $A^{*}$) and the M87$^{*}$ galaxy. This gives insight about if a supermassive BH can bear electric or magnetic fields of high intensities.

\begin{figure}[ht]
    \centering
    \includegraphics[width=0.8\textwidth]{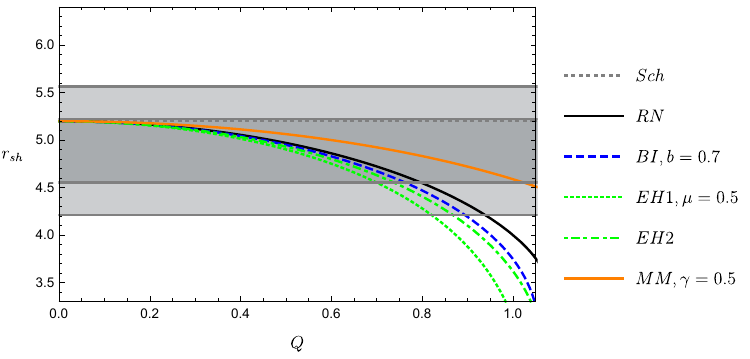}
    \caption{\footnotesize The shadow radii of the NLED BHs as a function of the BH charge are confronted with the shadow of the Sgr $A^{*}$ BH; the dark gray and light gray bands correspond to the $1\sigma$ and $2\sigma$, respectively. The dotted gray line indicates the radius of the Schwarzschild BH $r_{sh,Schw}= 5.196 M$, which does not depend on the charge; the black line is the RN BH, the dotted and dot-dashed lines in green are the two shadows obtained for the EH BH; the dashed blue line is the BI BH shadow; and the solid line in orange indicates the ModMax BH one. We are taking $M=1$.}
    \label{fig:compshadowbhs2}
\end{figure}

Fig. \ref{fig:compshadowbhs2} shows the shadow radii for the BHs in consideration as a function of the charge $Q$, contrasted with the $\sigma$ intervals given by the EHT for Sgr $A^{*}$ \cite{Vagnozzi2023}. As the charge increases the radius of the shadow decreases. 

The radius of the shadow for BI and the second solution of EH behave similarly for values of the nonlinear parameters $b=0.7, \mu=0.5$; for $ Q/M \le 0.2$ both BHs are in excellent agreement with the shadow of Sgr $A^{*}$ and up to $Q/M \le 0.7$ their shadows remain in the $1 \sigma$ band of Sgr $A^{*}$. The ModMax BH is the one that allows the largest charge/mass ratio $Q/M$ keeping its shadow radius in the $1\sigma$ region for $\gamma = 0.5$.

\begin{figure}[ht]
    \centering
    \includegraphics[width=0.8\textwidth]{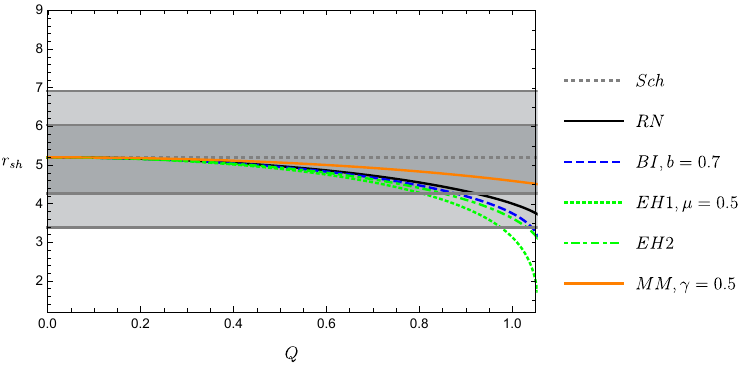}
    \caption{\footnotesize The shadow radii of the NLED BHs as a function of the BH charge $Q$ are confronted with the shadow of the  M87$^{*}$ BH; the dark gray and light gray bands correspond to the $1\sigma$ and $2\sigma$, respectively. The dotted gray line indicates the radius of the Schwarzschild BH which does not depend on the charge; the black line is the RN BH; the dotted and dot-dashed lines in green are the two shadows obtained for the EH BH; the dashed blue line is the BI BH shadow; and the solid line in orange indicates the ModMax BH one.  We are taking $M=1$.}
    \label{fig:compshadowbhsM87}
\end{figure}

In Fig. \ref{fig:compshadowbhsM87} the shadow radii for the NLED BHs is contrasted with the observations of  M87$^*$ \cite{Chakhchi2024,EHT_M87}. Note that the $1\sigma$ constraint reaches values above the radius for the Schwarzschild BH, nonetheless, the BHs in consideration do not exceed the Schwarzschild radius. 
The confidence intervals of M87$^*$  are less restrictive than the ones for Sgr $A^{*}$; which allows larger values of $Q/M$ BH in the $1 \sigma$ band.
For $Q/M \le 0.35$ EH and BI BHs are in excellent agreement with the shadow of M87$^*$and up to $ Q/M \le 0.8$ their shadows remain in the $1 \sigma$ band. While the ModMax BH remains in the $1 \sigma$ band for  $Q/M \le 1$ for $\gamma = 0.5$.

\begin{figure}[ht]
    \centering
    \includegraphics[width=0.8\textwidth]{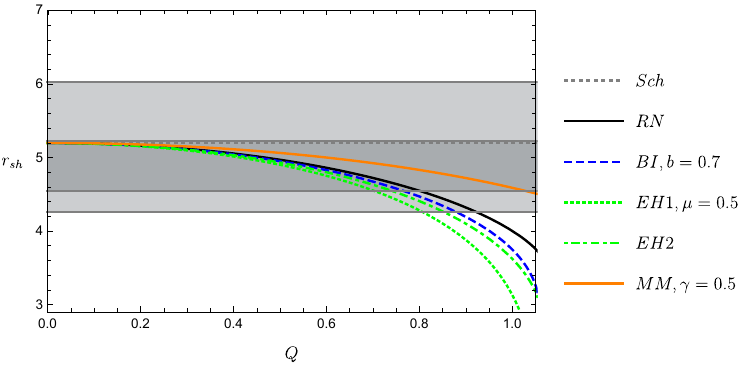}
    \caption{\footnotesize The shadow radii of the considered  NLED BHs as a function of the BH charge, $Q$,  are shown in the $1 \sigma$ band of the shadow observations for the Sgr $A^{*}$  (dark gray band) and M87$^{*}$ (light gray band) BHs.  The dotted gray line indicates the radius of the Schwarzschild BH which does not depend on the charge; the black line is the RN BH; the dotted and dot-dashed lines in green are the two shadows obtained for the EH BH; the dashed blue line is the BI BH shadow; and the solid line in orange indicates the ModMax BH one. We are taking $M=1$.}
    \label{fig:compshadowbhsSgrAM87}
\end{figure}

In Fig. \ref{fig:compshadowbhsSgrAM87} we compare the $1\sigma$ bands for the SgrA$^{*}$ and M87$^{*}$ being the light gray the $1\sigma$ band  for M87$^{*}$ and the dark gray the one for SgrA$^{*}$. As noted before, the confidence intervals for M87$^{*}$ include the ones for SgrA$^{*}$.

\begin{figure}[t]
\centering
\includegraphics[width=0.8\textwidth]{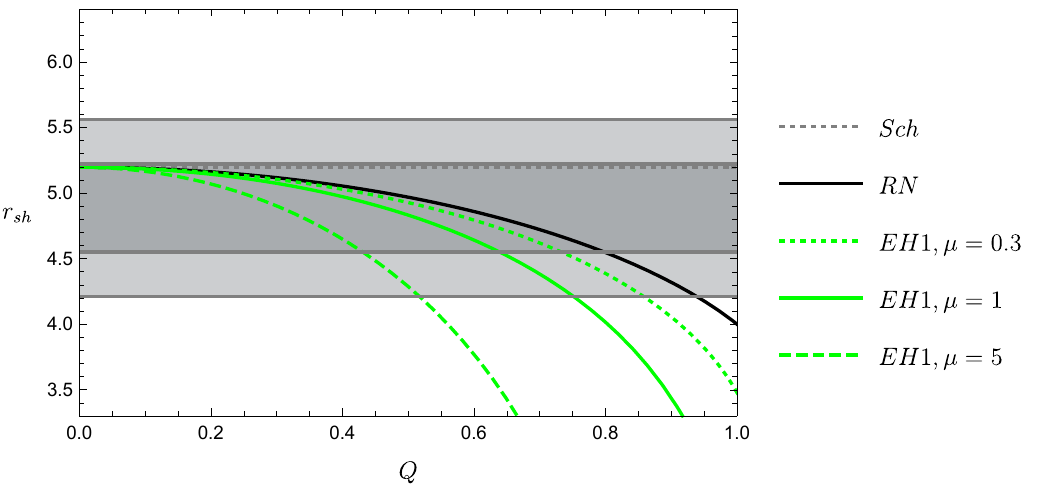} 
\includegraphics[width=0.8\textwidth]{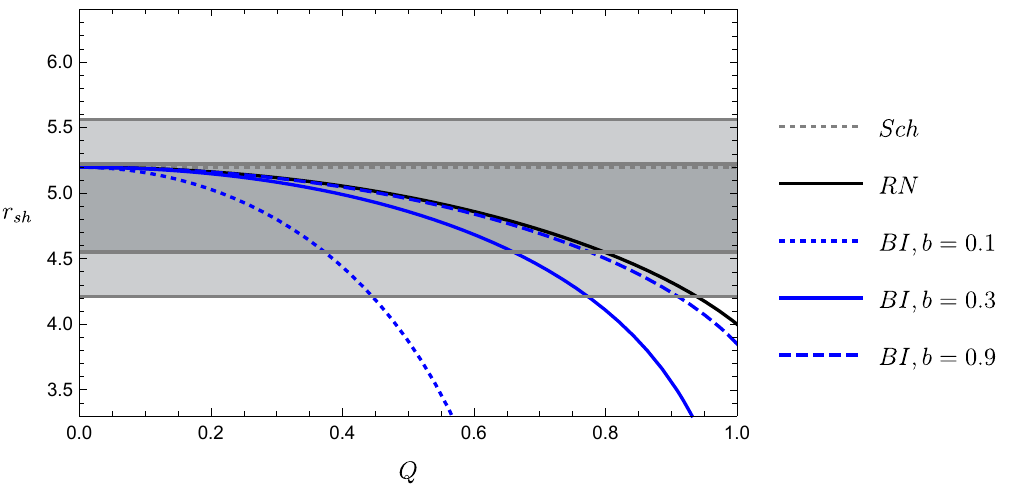}
\includegraphics[width=0.8\textwidth]{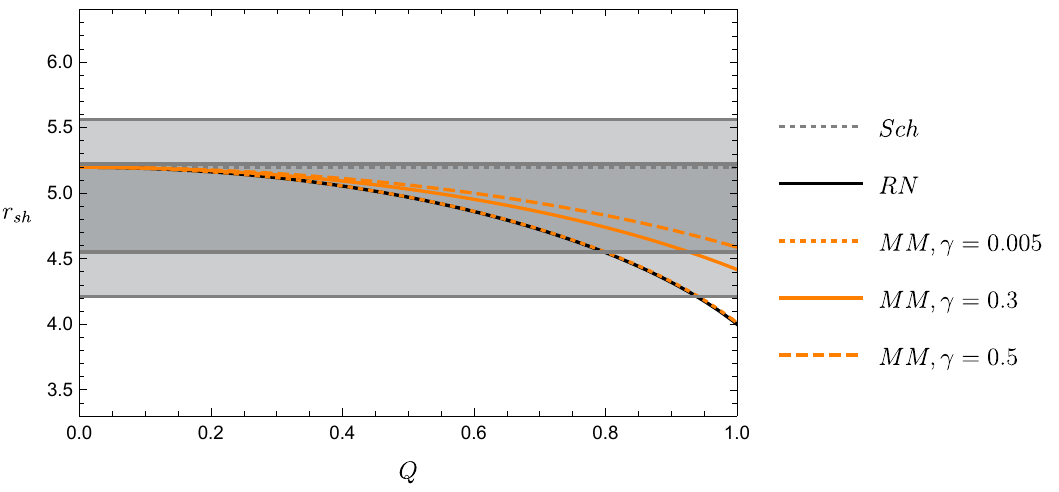}        
\caption{\footnotesize The shadow radii of the EH1, BI,  MM BHs and RN as a function of the BH charge are confronted with the observational shadow of the Sgr $A^{*}$ BH;
the dark gray and light ray bands correspond to the $1\sigma$ and $2\sigma$, respectively.  The dotted gray line indicates the radius of the Schwarzschild BH, and the black line is the RN BH. The green lines are the radius of the EH1 BH for different values of the nonlinear parameter $\mu$ (upper panel); the blue lines are the radius of the BI BH for different values of the nonlinear parameter $b$ (middle panel); the orange lines are the radius of the MM BH for different values of the nonlinear parameter $\gamma$ (lower panel).  We are taking $M=1$. }
    \label{fig:compshadow_all}
\end{figure}

Since tighter confidence intervals come from SgrA$^{*}$, 
in Fig. \ref{fig:compshadow_all} we confront the shadow radii of EH1, BI and MM BHs for different values of their respective nonlinear parameters $\mu$, $b$ and $\gamma$. For the EH BH, we only plot the first solution as the behaviors are similar; as the value of $\mu$ increases, the allowed values of the charge that fall in the $1\sigma$ and $2\sigma$ regions, are more restricted, however, $\mu=0.3$ admits charge-mass ratios of $0.8 \le Q/M$ in the $1 \sigma$ interval and $ 0.9 \le Q/M$ in the $2 \sigma$ interval. 

For the BI BH increasing $b$ the radius of the shadow tends to the one of the RN BH, as expected, since in the limit $b\rightarrow \infty$ Maxwell electrodynamics is recovered. It is important to highlight that Figure \ref{fig:compshadow_all} shows that BI and EH nonlinear theories do not allow higher values of the charge in agreement with the observations; then, if one looks for the existence of a charged black hole, neither BI or EH NLEDs are the theories that would help to achieve that, at least in the range of supermassive black holes; however, up to a relation charge mass $Q/M \approx 0.3$ charged BHs are not ruled out for any of the examined NLEDs.

In contrast, for the MM BH, if $\gamma\rightarrow 0$, we recover the radius of the shadow of the RN BH, and as $\gamma$ increases higher values of the BH charge are admitted; for $\gamma =0.5$ the whole range of $Q/M \le 1$ is in the $1 \sigma$ band.
Moreover, the MM BH fits better the shadow observations than RN.

\textcolor{black}{It is also interesting to find out how large a charge of $Q=0.3 M$ is in physical units, i.e. Coulombs, and if it is acceptable to consider that the supermassive BH Sgr $A^{*}$ or $M87^{*}$ could bear this electric charge excess. Considering that the order of magnitude of the allowed charge for RN is very close to the ones for the NLED BHs, and recalling that the RN-BH charge cannot be arbitrarily large but the existence of the horizon constrains the charge to the  inequality $Q^2 \le M^2$; recovering the physical units in the charge constrain for RN-BH,
\begin{equation}\label{RN_ch}
\frac{GQ^2}{c^4} \le   \frac{G^2M^2}{c^4},    
\end{equation}
that  reduces to
\begin{equation}
Q \le \sqrt{G}M = 7.74 \times 10^{-14} \frac{ \rm C}{\rm gr}M,    
\end{equation}
where C stands for Coulombs and the mass should be given in grams,
such that for one solar mass, $M_{\odot}= 1.9 \times 10^{33}$gr, 
then
\begin{equation}
Q \le  1.47 \times 10^{20} \left[ \frac{M}{M_{\odot}} \right] {\rm C},    
\end{equation}
where the mass $M$ should be given in solar masses. Substituting then for Sgr $A^{*}$ $M= 4 \times 10^{6} M_{\odot} $  and for  $M87^{*}$ $M= 2.4 \times 10^{9} M_{\odot} $, we have the constraint given by $Q \le 1.7 \times 10^{25}$C for Sgr $A^{*}$ and $Q \le 1.06 \times 10^{28}$C for $M87^{*}$ and for $Q \le 0.3 M$ the bounds are reduced to $Q \le 5.1 \times 10^{24}$C for Sgr $A^{*}$ and $Q \le 3.18 \times 10^{27}$C for $M87^{*}$. Particularly we should  compare this constraint with the bounds established for Sgr $A^{*}$ in \cite{Zajaček_2019}, where different constraints to the electric charge of Sgr $A^{*}$ are examined, coming from different effects, like the mass imbalance between electron-proton in fully ionized  plasmas, which yield a constraint of $10^{8}$C. Other processes  related to the magnetic field in rotation give a bound of $10^{15}$C; and the bound set for the RN extreme BH, $Q=M$, that corresponds to the saturation of Eq. (\ref{RN_ch}).
 Since  the charge value  of the extreme RN BH is at least ten orders of magnitude  larger  than the constraints in \cite{Zajaček_2019},  there is not influence on neutral particles trajectories, however, the dynamics of charged particles can be modified even by small amounts of charge.}
\section{Conclusions}
\label{Sec:Conclusions}

We considered three nonlinear electrodynamics (NLED) theories: Born-Infeld (BI), Euler-Heisenberg (EH), and modified Maxwell (ModMax, MM) to study light propagation in a strong magnetic (electric)  background; the effect of these intense fields in light propagation is the slowing down of the phase velocity. Another effect that arises due to the high intensities of the fields is birefringence in vacuum. For the NLED theories in consideration, only the BI theory does not present birefringence; and in the case of the ModMax theory, as a consequence of the conformal symmetry, one of the light trajectories coincides with the null geodesics of the Minkowski space.  The NLED theory that diminishes the phase velocity the most is the EH theory and the effect is enhanced in proportion to the magnetic (electric) component that is orthogonal to the propagation direction. 
 Our treatment is valid for very strong fields; if we consider, for instance, that $B^2/b^2 \approx 10^{-2}$, then the BI field is of the order of $10^{11}$ Tesla that is one hundred times the critical Schwinger field or  $B_{\rm cr} \approx 10^{9}$ Tesla. It is important to note that for fields of the orders $B^2/b^2\approx10^{-2}$ and $\mu B^2\approx 10^{-2}$ the diminishing in the velocities for BI and EH is qualitatively similar, as illustrated in  Figure \ref{fig:compEHBIMM1}. The plots that serve the comparison are in terms of dimensionless parameters $\mu B^2$ for EH and $B^2/b^2$ for BI. Still, we should keep in mind that the critical fields of both nonlinear theories are different by two orders of magnitude, being the BI theory the one with the larger critical field. 
 
 Additionally, the comparison between the phase velocities is useful to set bounds on the ModMax nonlinear parameter $\gamma$ since so far it is not constrained apart from the restriction of being positive that is derived from the energy conditions in the ModMax BH. BI constrains $\gamma$ to the interval  $0 < \gamma < 0.549306$; the larger bound for BI fields $ B \approx b$ with $b = 10^{11}$Tesla  ($b = 10^{20}$ Volt/m)  being the maximum attainable electromagnetic BI field. Then
we derive that the EH theory is more stringent constraining  $\gamma$ to the interval $ 0.0001 <\gamma < 0.00018$ for fields of the order of $B_{cr} \sim 10^{9}$ Tesla  ( $E_{cr}=10^{18}$ Volt/m ).  
In the second part of the paper we contemplate light propagation in the vicinity of NLED BHs as the background metric and calculate the corresponding shadows 
to compare the NLED effects with the ones for the linear solution of the Einstein-Maxwell equations, the Reissner-Nordstrom (RN) metric and, as a reference, we include in the comparison the Schwarzschild BH. Then these shadows are
confronted with the observational shadows from Sagittarius $A^{\ast}$ and M87$^{*}$.  The situation is different for each electrodynamics.  Due to birefringence, EH presents two shadows and BI only one. For ModMax one of the effective metrics is the background metric as a consequence of the conformal invariance.

We can determine ranges for the nonlinear parameters $\mu$ for EH, $b$ for BI, and $\gamma$ for ModMax, for their shadows to be in agreement with the evidence of Sagittarius $A^{*}$ and M87$^{*}$.
 If one is looking for a charged BH, the most effective theory is ModMax, which coupled with Einstein equations results in a metric function similar to the one for RN but with a screened charge; the ModMax BH allows  values of the BH charge-mass $Q/M \le 1$ still in the $1 \sigma$ confidence interval with the observations of Sgr. $A^{*}$ and M87$^{*}$. Meanwhile, by increasing the BI and EH nonlinear parameters, we are not able to reach higher values of the BH charge that keep the shadow in the $1\sigma$ interval of Sgr.$A^*$ that sets the most stringent astrophysical limits for the NLED parameters. 

 In summary, up to a relation charge mass $Q/M \le 0.3$  astrophysical charged BH are not ruled out for any of the examined NLEDs; their shadows remain in the $1\sigma$ interval of Sgr.$A^*$.  However neither BI-BH or EH-BH point to a possible astrophysical supermassive charged BH with $Q/M > 0.3$. In contrast, the Mod Max BH shadow remains in the $1 \sigma$ interval for BH charge-mass  $0  \le Q/M \le 1$, fitting data even better than the Einstein-Maxwell solution RN.

\acknowledgments
EGH acknowledges financial support by Conahcyt-Mexico through the Ph. D.  scholarship No. 761878.
The work of AM has been sponsored by Conahcyt-Mexico through the Posdoc Project I1200/311/2023. 
NB acknowledges partial support by Conahcyt-Mexico 
project CBF2023-2024-811.


\bibliographystyle{JHEP}
\bibliography{biblio.bib}

\providecommand{\href}[2]{#2}\begingroup\raggedright\begin{thebibliography}{10}

\bibitem{Dunne2012}
G.V.~Dunne, \emph{{The Heisenberg-Euler effective action: 75 Years on}},
  \href{https://doi.org/10.1142/S2010194512007222}{\emph{International Journal
  of Modern Physics: Conference Series} {\bfseries {\bf 14}} (2012) 42}.

\bibitem{Denisov2014EffectsON}
D.M.~Denisov, \emph{Effects of nonlinear electrodynamics in the magnetic field
  of a pulsar}, {\emph{Canadian Journal of Physics} {\bfseries 92} (2014)
  1453}.

\bibitem{Abishev:2016xxa}
M.~Abishev, Y.~Aimuratov, Y.~Aldabergenov, N.~Beissen, Z.~Bakytzhan and
  M.~Takibayeva, \emph{{Some astrophysical effects of nonlinear vacuum
  electrodynamics in the magnetosphere of a pulsar}},
  \href{https://doi.org/10.1016/j.astropartphys.2015.04.010}{\emph{Astropart.
  Phys.} {\bfseries 73} (2016) 8}.

\bibitem{Kaspi:2017fwg}
V.M.~Kaspi and A.~Beloborodov, \emph{{Magnetars}},
  \href{https://doi.org/10.1146/annurev-astro-081915-023329}{\emph{Ann. Rev.
  Astron. Astrophys.} {\bfseries 55} (2017) 261}
  [\href{https://arxiv.org/abs/1703.00068}{{\ttfamily 1703.00068}}].

\bibitem{PhysRevD.71.063002}
V.I.~Denisov and S.I.~Svertilov, \emph{Nonlinear electromagnetic and
  gravitational actions of neutron star fields on electromagnetic wave
  propagation}, \href{https://doi.org/10.1103/PhysRevD.71.063002}{\emph{Phys.
  Rev. D} {\bfseries 71} (2005) 063002}.

\bibitem{MosqueraCuesta:2003dh}
H.J.~Mosquera~Cuesta and J.M.~Salim, \emph{{Nonlinear electrodynamics and the
  surface redshift of pulsars}},
  \href{https://doi.org/10.1086/378686}{\emph{Astrophys. J.} {\bfseries 608}
  (2004) 925} [\href{https://arxiv.org/abs/astro-ph/0307513}{{\ttfamily
  astro-ph/0307513}}].

\bibitem{2014ApJS..212....6O}
S.A.~{Olausen} and V.M.~{Kaspi}, \emph{{The McGill Magnetar Catalog}},
  \href{https://doi.org/10.1088/0067-0049/212/1/6}{\emph{ApJS} {\bfseries 212}
  (2014) 6} [\href{https://arxiv.org/abs/1309.4167}{{\ttfamily 1309.4167}}].

\bibitem{Baring2008}
M.G.~Baring, \emph{{Photon splitting and pair conversion in strong magnetic
  fields}},  in \emph{AIP Conference Proceedings}, vol.~{{\bf 1051 }},
  pp.~53--64, American Institute of Physics, 2008,
  \href{https://doi.org/10.1063/1.3020681}{DOI}.

\bibitem{Mignani2017}
R.P.~Mignani, V.~Testa, D.~Gonz{\'a}lez-Caniulef, R.~Taverna, R.~Turolla,
  S.~Zane et~al., \emph{{Evidence for vacuum birefringence from the first
  optical polarimetry measurement of the isolated neutron star RX J1856. 5-
  3754}}, \href{https://doi.org/10.1093/mnras/stw2798}{\emph{Monthly Notices of
  the Royal Astronomical Society} {\bfseries {\bf 465}} (2016) 2798}.

\bibitem{Denisov2017}
V.~Denisov, V.~Sokolov and S.~Svertilov, \emph{Vacuum nonlinear electrodynamic
  polarization effects in hard emission of pulsars and magnetars},
  \href{https://doi.org/10.1088/1475-7516/2017/09/004}{\emph{Journal of
  Cosmology and Astroparticle Physics} {\bfseries 2017} (2017) 004}.

\bibitem{Zavattini2008}
{\scshape PVLAS} collaboration, \emph{{New PVLAS results and limits on
  magnetically induced optical rotation and ellipticity in vacuum}},
  \href{https://doi.org/10.1103/PhysRevD.77.032006}{\emph{Physical Review D}
  {\bfseries {\bf 77}} (2008) 032006}.

\bibitem{Zavattini2012}
G.~Zavattini, U.~Gastaldi, R.~Pengo, G.~Ruoso, F.D.~Valle and E.~Milotti,
  \emph{{Measuring the magnetic birefringence of vacuum: the PVLAS
  experiment}},
  \href{https://doi.org/10.1142/S0217751X12600172}{\emph{International Journal
  of Modern Physics A} {\bfseries {\bf 27}} (2012) 1260017}.

\bibitem{Ejlli:2020yhk}
A.~Ejlli, F.~Della~Valle, U.~Gastaldi, G.~Messineo, R.~Pengo, G.~Ruoso et~al.,
  \emph{{The PVLAS experiment: A 25 year effort to measure vacuum magnetic
  birefringence}},
  \href{https://doi.org/10.1016/j.physrep.2020.06.001}{\emph{Phys. Rept.}
  {\bfseries 871} (2020) 1} [\href{https://arxiv.org/abs/2005.12913}{{\ttfamily
  2005.12913}}].

\bibitem{Cameron1993}
R.~Cameron, G.~Cantatore, A.C.~Melissinos, G.~Ruoso, Y.~Semertzidis,
  H.J.~Halama et~al., \emph{{Search for nearly massless, weakly coupled
  particles by optical techniques}},
  \href{https://doi.org/10.1103/PhysRevD.47.3707}{\emph{Physical Review D}
  {\bfseries {\bf 47}} (1993) 3707}.

\bibitem{Cadene:2013bva}
A.~Cad\`ene, P.~Berceau, M.~Fouch\'e, R.~Battesti and C.~Rizzo, \emph{{Vacuum
  magnetic linear birefringence using pulsed fields: status of the BMV
  experiment}}, \href{https://doi.org/10.1140/epjd/e2013-40725-9}{\emph{Eur.
  Phys. J. D} {\bfseries 68} (2014) 16}
  [\href{https://arxiv.org/abs/1302.5389}{{\ttfamily 1302.5389}}].

\bibitem{Tommasini:2009nh}
D.~Tommasini, A.~Ferrando, H.~Michinel and M.~Seco, \emph{{Precision tests of
  QED and non-standard models by searching photon-photon scattering in vacuum
  with high power lasers}},
  \href{https://doi.org/10.1088/1126-6708/2009/11/043}{\emph{JHEP} {\bfseries
  11} (2009) 043} [\href{https://arxiv.org/abs/0909.4663}{{\ttfamily
  0909.4663}}].

\bibitem{Karbstein:2021otv}
F.~Karbstein, \emph{{Vacuum Birefringence at the Gamma Factory}},
  \href{https://doi.org/10.1002/andp.202100137}{\emph{Annalen Phys.} {\bfseries
  534} (2022) 2100137} [\href{https://arxiv.org/abs/2106.06359}{{\ttfamily
  2106.06359}}].

\bibitem{Ahmadiniaz2020}
N.~Ahmadiniaz, T.E.~Cowan, R.~Sauerbrey, U.~Schramm, H.-P.~Schlenvoigt and
  R.~Sch{\"u}tzhold, \emph{Heisenberg limit for detecting vacuum
  birefringence},
  \href{https://doi.org/10.1103/PhysRevD.101.116019}{\emph{Physical Review D}
  {\bfseries 101} (2020) 116019}.

\bibitem{fouche2016limits}
M.~Fouch{\'e}, R.~Battesti and C.~Rizzo, \emph{{Limits on nonlinear
  electrodynamics}},
  \href{https://doi.org/10.1103/PhysRevD.93.093020}{\emph{Physical Review D}
  {\bfseries {\bf 93}} (2016) 093020}.

\bibitem{Atlas2017}
{\scshape ATLAS} collaboration, \emph{{Evidence for light-by-light scattering
  in heavy-ion collisions with the ATLAS detector at the LHC}},
  \href{https://doi.org/10.1038/nphys4208}{\emph{Nature Phys.} {\bfseries 13}
  (2017) 852} [\href{https://arxiv.org/abs/1702.01625}{{\ttfamily
  1702.01625}}].

\bibitem{Enterria2013}
D.~d'Enterria and G.G.~da~Silveira, \emph{Observing light-by-light scattering
  at the large hadron collider},
  \href{https://doi.org/10.1103/PhysRevLett.111.080405}{\emph{Phys. Rev. Lett.}
  {\bfseries 111} (2013) 080405}.

\bibitem{Fedotov2023}
A.~Fedotov, A.~Ilderton, F.~Karbstein, B.~King, D.~Seipt, H.~Taya et~al.,
  \emph{{Advances in QED with intense background fields}},
  \href{https://doi.org/10.1016/j.physrep.2023.01.003}{\emph{Physics Reports}
  {\bfseries {\bf 1010}} (2023) 1}.

\bibitem{Schoeffel2021}
L.~Schoeffel, C.~Baldenegro, H.~Hamdaoui, S.~Hassani, C.~Royon and M.~Saimpert,
  \emph{Photon--photon physics at the lhc and laser beam experiments, present
  and future}, \href{https://doi.org/10.1016/j.ppnp.2021.103889}{\emph{Progress
  in Particle and Nuclear Physics} {\bfseries 120} (2021) 103889}.

\bibitem{tzenov2020dispersion}
S.I.~Tzenov, K.M.~Spohr and K.A.~Tanaka, \emph{{Dispersion properties,
  nonlinear waves and birefringence in classical nonlinear electrodynamics}},
  \href{https://doi.org/10.1088/2399-6528/ab72c7}{\emph{Journal of Physics
  Communications} {\bfseries {\bf 4}} (2020) 025006}.

\bibitem{BI1934}
M.~Born and L.~Infeld, \emph{{Foundations of the new field theory}},
  \href{https://doi.org/10.1098/rspa.1934.0059}{\emph{Proceedings of the Royal
  Society of London. Series A, Containing Papers of a Mathematical and Physical
  Character} {\bfseries {\bf 144}} (1934) 425}.

\bibitem{Rebhan2017}
A.~Rebhan and G.~Turk, \emph{{Polarization effects in light-by-light
  scattering: Euler--Heisenberg versus Born--Infeld}},
  \href{https://doi.org/10.1142/S0217751X17500531}{\emph{International Journal
  of Modern Physics A} {\bfseries {\bf 32}} (2017) 1750053}.

\bibitem{Kadlecova2024}
H.~Kadlecová, \emph{{On the absence of shock waves and vacuum birefringence in
  Born–Infeld electrodynamics}},
  \href{https://doi.org/10.1063/5.0150790}{\emph{Journal of Mathematical
  Physics} {\bfseries {\bf 65}} (2024) 012302}.

\bibitem{Schellstede2015}
G.O.~Schellstede, V.~Perlick and C.~L\"ammerzahl, \emph{{Testing non-linear
  vacuum electrodynamics with Michelson interferometry}},
  \href{https://doi.org/10.1103/PhysRevD.92.025039}{\emph{Phys. Rev. D}
  {\bfseries 92} (2015) 025039}
  [\href{https://arxiv.org/abs/1504.03159}{{\ttfamily 1504.03159}}].

\bibitem{Fradkin1985}
E.S.~Fradkin and A.A.~Tseytlin, \emph{{Nonlinear Electrodynamics from Quantized
  Strings}}, \href{https://doi.org/10.1016/0370-2693(85)90205-9}{\emph{Phys.
  Lett. B} {\bfseries 163} (1985) 123}.

\bibitem{Babaei-Aghbolagh:2022uij}
H.~Babaei-Aghbolagh, K.B.~Velni, D.M.~Yekta and H.~Mohammadzadeh,
  \emph{{Emergence of non-linear electrodynamic theories from
  TT\textasciimacron{}-like deformations}},
  \href{https://doi.org/10.1016/j.physletb.2022.137079}{\emph{Phys. Lett. B}
  {\bfseries 829} (2022) 137079}
  [\href{https://arxiv.org/abs/2202.11156}{{\ttfamily 2202.11156}}].

\bibitem{EH}
E.~Heisenberg and H.~Euler, \emph{{Folgerungen aus der Diracshen Theorie des
  Positrons}}, \href{https://doi.org/10.1007/BF01343663}{\emph{Zeitschrift fur
  Physik} {\bfseries {\bf 98}} (1936) 714}.

\bibitem{Dunne2021}
G.V.~Dunne and Z.~Harris, \emph{{Higher-loop Euler-Heisenberg transseries
  structure}},
  \href{https://doi.org/10.1103/PhysRevD.103.065015}{\emph{Physical Review D}
  {\bfseries {\bf 103}} (2021) 065015}.

\bibitem{Sasorov2021}
P.V.~Sasorov, F.~Pegoraro, T.Z.~Esirkepov and S.V.~Bulanov, \emph{{Generation
  of high order harmonics in Heisenberg--Euler electrodynamics}},
  \href{https://doi.org/10.1088/1367-2630/ac28cb}{\emph{New Journal of Physics}
  {\bfseries {\bf 23}} (2021) 105003}.

\bibitem{Bandos2020}
I.~Bandos, K.~Lechner, D.~Sorokin and P.K.~Townsend, \emph{{Nonlinear
  duality-invariant conformal extension of Maxwell's equations}},
  \href{https://doi.org/10.1103/PhysRevD.102.121703}{\emph{Physical Review D}
  {\bfseries {\bf 102}} (2020) 121703}.

\bibitem{Banerjee2022}
A.~Banerjee and A.~Mehra, \emph{{Maximally symmetric nonlinear extension of
  electrodynamics with Galilean conformal symmetries}},
  \href{https://doi.org/10.1103/PhysRevD.106.085005}{\emph{Physical Review D}
  {\bfseries {\bf 106}} (2022) 085005}.

\bibitem{Neves2023}
M.J.~Neves, P.~Gaete, L.P.R.~Ospedal and J.A.~Helay\"el-Neto,
  \emph{{Considerations on the modified Maxwell electrodynamics in the presence
  of an electric and magnetic background}},
  \href{https://doi.org/10.1103/PhysRevD.107.075019}{\emph{Physical Review D}
  {\bfseries {\bf 107}} (2023) 075019}.

\bibitem{Escobar2022}
C.A.~Escobar and R.~Linares, \emph{{Spontaneous symmetry breaking in models
  with second-class constraints}},
  \href{https://doi.org/10.1103/PhysRevD.106.036027}{\emph{Physical Review D}
  {\bfseries {\bf 106}} (2022) 036027}.

\bibitem{Bandos2021p}
I.~Bandos, K.~Lechner, D.~Sorokin and et~al., \emph{{On p-form gauge theories
  and their conformal limits}},
  \href{https://doi.org/10.1007/JHEP03(2021)022}{\emph{Journal of High Energy
  Physics} {\bfseries {\bf 22}} (2021) }.

\bibitem{BandosSusy}
I.~Bandos, L.~K., D.~Sorokin and T.P.~K., \emph{{ModMax meets SUSY}},
  \href{https://doi.org/10.1007/JHEP10(2021)031}{\emph{Journal of High Energy
  Physics} {\bfseries {\bf 2021}} (2021) 1}.

\bibitem{Barrientos2022}
J.~Barrientos, A.~Cisterna, D.~Kubizňák and J.~Oliva, \emph{{Accelerated
  black holes beyond Maxwell's electrodynamics}},
  \href{https://doi.org/10.1016/j.physletb.2022.137447}{\emph{Physics Letters
  B} {\bfseries {\bf 834}} (2022) 137447}.

\bibitem{Bordo2021}
A.~Ballon~Bordo, D.~Kubizňák and T.R.~Perche, \emph{{Taub-NUT solutions in
  conformal electrodynamics}},
  \href{https://doi.org/10.1016/j.physletb.2021.136312}{\emph{Physics Letters
  B} {\bfseries {\bf 817}} (2021) 136312}.

\bibitem{Bokulic2021}
A.~Bokuli\'{c}, I.~Smoli\'{c} and T.~Juri\'{c}, \emph{{Black hole
  thermodynamics in the presence of nonlinear electromagnetic fields}},
  \href{https://doi.org/10.1103/PhysRevD.103.124059}{\emph{Physical Review D}
  {\bfseries {\bf 103}} (2021) 124059}.

\bibitem{Pantig2022}
R.C.~Pantig, L.~Mastrototaro, G.~Lambiase and A.~Övgün, \emph{{Shadow,
  lensing, quasinormal modes, greybody bounds and neutrino propagation by
  dyonic ModMax black holes}},
  \href{https://doi.org/10.1140/epjc/s10052-022-11125-y}{\emph{The European
  Physical Journal C} {\bfseries {\bf 82}} (2022) 1155}.

\bibitem{Babaei2022}
H.~Babaei-Aghbolagh, K.B.~Velni, D.M.~Yekta and H.~Mohammadzadeh,
  \emph{{Marginal $T\overline{T}$-like deformation and modified Maxwell
  theories in two dimensions}},
  \href{https://doi.org/10.1103/PhysRevD.106.086022}{\emph{Physical Review D}
  {\bfseries {\bf 106}} (2022) 086022}.

\bibitem{Boillat1970}
G.~Boillat, \emph{{Nonlinear Electrodynamics: Lagrangians and Equations of
  Motion}}, \href{https://doi.org/10.1063/1.1665231}{\emph{Journal of
  Mathematical Physics} {\bfseries {\bf 11}} (2003) 941}.

\bibitem{Bialynicka1970}
Z.~Bialynicka-Birula and I.~Bialynicki-Birula, \emph{{Nonlinear effects in
  quantum electrodynamics. Photon propagation and photon splitting in an
  external field}},
  \href{https://doi.org/10.1103/PhysRevD.2.2341}{\emph{Physical Review D}
  {\bfseries {\bf 2}} (1970) 2341}.

\bibitem{Pleban}
S.~Alarcón~Gutiérrez, A.L.~Dudley and J.F.~Plebanski, \emph{{Signals and
  discontinuities in general relativistic nonlinear electrodynamics}},
  \href{https://doi.org/10.1063/1.524874}{\emph{Journal of Mathematical
  Physics} {\bfseries {\bf 22}} (1981) 2835}.

\bibitem{Novello2000}
M.~Novello, V.A.~De~Lorenci, J.M.~Salim and R.~Klippert, \emph{{Geometrical
  aspects of light propagation in nonlinear electrodynamics}},
  \href{https://doi.org/10.1103/PhysRevD.61.045001}{\emph{Physical Review D}
  {\bfseries {\bf 61}} (2000) 045001}.

\bibitem{Novello2000b}
V.~{De Lorenci}, R.~Klippert, M.~Novello and J.~Salim, \emph{{Light propagation
  in non-linear electrodynamics}},
  \href{https://doi.org/10.1016/S0370-2693(00)00522-0}{\emph{Physics Letters B}
  {\bfseries {\bf 482}} (2000) 134}.

\bibitem{Obukov2002}
Y.N.~Obukhov and G.F.~Rubilar, \emph{{Fresnel analysis of wave propagation in
  nonlinear electrodynamics}},
  \href{https://doi.org/10.1103/PhysRevD.66.024042}{\emph{Physical Review D}
  {\bfseries {\bf 66}} (2002) 024042}.

\bibitem{Goulart2009}
E.~Goulart and S.E.~Perez-Bergliaffa, \emph{{A classification of the effective
  metric in nonlinear electrodynamics}},
  \href{https://doi.org/10.1088/0264-9381/26/13/135015}{\emph{Classical and
  Quantum Gravity} {\bfseries {\bf 26}} (2009) 135015}.

\bibitem{Vagnozzi2023}
S.~Vagnozzi, R.~Roy, Y.~Tsai, L.~Visinelli, M.~Afrin, A.~Allahyari et~al.,
  \emph{{Horizon-scale tests of gravity theories and fundamental physics from
  the Event Horizon Telescope image of Sagittarius A}},
  \href{https://doi.org/10.1088/1361-6382/acd97b}{\emph{Classical and Quantum
  Gravity} {\bfseries {\bf 40}} (2022) }.

\bibitem{Chakhchi2024}
L.~Chakhchi, H.~{El Moumni} and K.~Masmar, \emph{{Signatures of the
  accelerating black holes with a cosmological constant from the Sgr A* and
  M87* shadow prospects}},
  \href{https://doi.org/https://doi.org/10.1016/j.dark.2024.101501}{\emph{Physics
  of the Dark Universe} {\bfseries 44} (2024) 101501}.

\bibitem{EHT_M87}
{\scshape EHT} collaboration, \emph{Constraints on black-hole charges with the
  2017 eht observations of m87*},
  \href{https://doi.org/10.1103/PhysRevD.103.104047}{\emph{Phys. Rev. D}
  {\bfseries 103} (2021) 104047}.

\bibitem{Allahyari:2019jqz}
A.~Allahyari, M.~Khodadi, S.~Vagnozzi and D.F.~Mota, \emph{{Magnetically
  charged black holes from non-linear electrodynamics and the Event Horizon
  Telescope}}, \href{https://doi.org/10.1088/1475-7516/2020/02/003}{\emph{JCAP}
  {\bfseries 02} (2020) 003}
  [\href{https://arxiv.org/abs/1912.08231}{{\ttfamily 1912.08231}}].

\bibitem{Ruffini2013}
R.~Ruffini, Y.-B.~Wu and S.-S.~Xue, \emph{{Einstein-Euler-Heisenberg Theory and
  charged black holes}},
  \href{https://doi.org/10.1103/PhysRevD.88.085004}{\emph{Phys. Rev. D}
  {\bfseries 88} (2013) 085004}
  [\href{https://arxiv.org/abs/1307.4951}{{\ttfamily 1307.4951}}].

\bibitem{breton2002geodesic}
N.~Bret{\'o}n, \emph{{Geodesic structure of the Born--Infeld black hole}},
  \href{https://doi.org/10.1088/0264-9381/19/4/301}{\emph{Classical and Quantum
  Gravity} {\bfseries {\bf 19}} (2002) 601}.

\bibitem{Maceda2020}
D.~Flores-Alfonso, B.A.~Gonz{\'a}lez-Morales, R.~Linares and M.~Maceda,
  \emph{{Black holes and gravitational waves sourced by non-linear duality
  rotation-invariant conformal electromagnetic matter}},
  \href{https://doi.org/10.1016/j.physletb.2020.136011}{\emph{Physics Letters
  B} {\bfseries {\bf 812}} (2021) 136011}.

\bibitem{Magos2020}
D.~Magos and N.~Breton, \emph{Thermodynamics of the euler-heisenberg-ads black
  hole}, \href{https://doi.org/10.1103/PhysRevD.102.084011}{\emph{Phys. Rev. D}
  {\bfseries 102} (2020) 084011}.

\bibitem{Gao:2021kvr}
C.~Gao, \emph{{Black holes with many horizons in the theories of nonlinear
  electrodynamics}},
  \href{https://doi.org/10.1103/PhysRevD.104.064038}{\emph{Phys. Rev. D}
  {\bfseries 104} (2021) 064038}
  [\href{https://arxiv.org/abs/2106.13486}{{\ttfamily 2106.13486}}].

\bibitem{Ruffini:1999kq}
R.~Ruffini, \emph{{The Dyadosphere of black holes and gamma-ray bursts}},
  \href{https://doi.org/10.1051/aas:1999331}{\emph{Astron. Astrophys. Suppl.
  Ser.} {\bfseries 138} (1999) 513}
  [\href{https://arxiv.org/abs/astro-ph/9905072}{{\ttfamily
  astro-ph/9905072}}].

\bibitem{Christodoulou:1971pcn}
D.~Christodoulou and R.~Ruffini, \emph{{Reversible transformations of a charged
  black hole}}, \href{https://doi.org/10.1103/PhysRevD.4.3552}{\emph{Phys. Rev.
  D} {\bfseries 4} (1971) 3552}.

\bibitem{TDamour1975}
T.~Damour and R.~Ruffini, \emph{Quantum electrodynamical effects in
  kerr-newmann geometries},
  \href{https://doi.org/10.1103/PhysRevLett.35.463}{\emph{Phys. Rev. Lett.}
  {\bfseries 35} (1975) 463}.

\bibitem{Ruffini:2009hg}
R.~Ruffini, G.~Vereshchagin and S.-S.~Xue, \emph{{Electron-positron pairs in
  physics and astrophysics: from heavy nuclei to black holes}},
  \href{https://doi.org/10.1016/j.physrep.2009.10.004}{\emph{Phys. Rept.}
  {\bfseries 487} (2010) 1} [\href{https://arxiv.org/abs/0910.0974}{{\ttfamily
  0910.0974}}].

\bibitem{Perlick2015}
V.~Perlick, O.Y.~Tsupko and G.S.~Bisnovatyi-Kogan, \emph{{Influence of a plasma
  on the shadow of a spherically symmetric black hole}},
  \href{https://doi.org/10.1103/PhysRevD.92.104031}{\emph{Physical Review D}
  {\bfseries {\bf 92}} (2015) 104031}.

\bibitem{Amaro2022}
D.~Amaro and A.~Mac{\'\i}as, \emph{{Exact lens equation for the
  Einstein-Euler-Heisenberg static black hole}},
  \href{https://doi.org/10.1103/PhysRevD.106.064010}{\emph{Physical Review D}
  {\bfseries {\bf 106}} (2022) 064010}.

\bibitem{Amaro2020}
D.~Amaro and A.~Mac{\'\i}as, \emph{{Geodesic structure of the Euler-Heisenberg
  static black hole}},
  \href{https://doi.org/10.1103/PhysRevD.102.104054}{\emph{Physical Review D}
  {\bfseries {\bf 102}} (2020) 104054}.

\bibitem{Perlick2018}
V.~Perlick, O.Y.~Tsupko and G.S.~Bisnovatyi-Kogan, \emph{{Black hole shadow in
  an expanding universe with a cosmological constant}},
  \href{https://doi.org/10.1103/PhysRevD.97.104062}{\emph{Physical Review D}
  {\bfseries {\bf 97}} (2018) 104062}.

\end{thebibliography}\endgroup





\end{document}